\documentclass[twocolumn,twocolappendix]{aastex631}  

\newcommand{\ha}{H$\alpha$}
\newcommand{\hb}{H$\beta$}
\newcommand{\paa}{Pa$\alpha$}
\newcommand{\pab}{Pa$\beta$}
\newcommand{\bra}{Br$\alpha$}
\newcommand{\brb}{Br$\beta$}

\usepackage{multirow}
\usepackage{amsmath}
\usepackage{color} 
\usepackage{makecell}
\graphicspath{{figures/}}     


\begin{document}

\title{A JWST View of Cosmic Star Formation from $z\sim0$ to 9 with Hydrogen Recombination Lines\\ in the COSMOS and GOODS-N fields}

\author[0000-0003-0964-7188]{Shuqi Fu}
\affiliation{Department of Astronomy, School of Physics, Peking University, Beijing 100871, People's Republic of China}
\affiliation{Kavli Institute for Astronomy and Astrophysics, Peking University, Beijing 100871, People's Republic of China}
\affiliation{Jodrell Bank Centre for Astrophysics, University of Manchester, Oxford Road, Manchester UK}

\author[0000-0003-4176-6486]{Linhua Jiang}
\affiliation{Department of Astronomy, School of Physics, Peking University, Beijing 100871, People's Republic of China}
\affiliation{Kavli Institute for Astronomy and Astrophysics, Peking University, Beijing 100871, People's Republic of China}

\author[0000-0003-1949-7638]{Christopher J. Conselice}
\affiliation{Jodrell Bank Centre for Astrophysics, University of Manchester, Oxford Road, Manchester UK}

\author[0000-0002-2420-5022]{Zijian Zhang}
\affiliation{Department of Astronomy, School of Physics, Peking University, Beijing 100871, People's Republic of China}
\affiliation{Kavli Institute for Astronomy and Astrophysics, Peking University, Beijing 100871, People's Republic of China}

\author[0009-0003-6747-2221]{Danyang Jiang}
\affiliation{Department of Astronomy, School of Physics, Peking University, Beijing 100871, People's Republic of China}
\affiliation{Kavli Institute for Astronomy and Astrophysics, Peking University, Beijing 100871, People's Republic of China}

\author[0000-0002-4765-1500]{Jie Chen}
\affiliation{Department of Astronomy, School of Physics, Peking University, Beijing 100871, People's Republic of China}
\affiliation{Kavli Institute for Astronomy and Astrophysics, Peking University, Beijing 100871, People's Republic of China}

\author[0009-0008-8375-8605]{Louis Quilley}
\affiliation{Jodrell Bank Centre for Astrophysics, University of Manchester}

\author[0000-0002-4622-6617]{Fengwu Sun}
\affiliation{Department of Astronomy, School of Science, Westlake University, Hangzhou, Zhejiang 310030, P.\ R.\ China}

\author[0000-0002-0322-6131]{Ronaldo Laishram}
\affiliation{National Astronomical Observatory of Japan, 2-21-1 Osawa, Mitaka, Tokyo 181-8588, Japan}

\author[0000-0003-1006-924X]{Alberto Traina}
\affiliation{Istituto Nazionale di Astrofisica (INAF) – Osservatorio di Astrofisica e Scienza dello Spazio (OAS), Via Gobetti 101, I-40129 Bologna, Italy}

\author[0009-0003-0031-0527]{Bahareh S. Salmasi}
\affiliation{School of Mathematical and Physical Sciences, Macquarie University, Sydney, NSW 2109, Australia}
\affiliation{Astrophysics and Space Technologies Research Centre, Macquarie University, Sydney, NSW 2109, Australia}

\author[0000-0003-0519-9445]{Duncan Austin}
\affiliation{Jodrell Bank Centre for Astrophysics, University of Manchester, Oxford Road, Manchester UK}

\begin{abstract}

We combine wide-area COSMOS-3D observations with deeper GOODS-N data from JWST/NIRCam F444W wide-field slitless spectroscopy to trace cosmic star formation over $z\sim0$--9.
The observations cover approximately $0.33~\mathrm{deg}^2$ in COSMOS and $60.9~\mathrm{arcmin}^2$ in GOODS-N, with typical $3\sigma$ integrated line-flux limits of $3.2\times10^{-18}$ and $1.04\times10^{-18}~\mathrm{erg\,s^{-1}\,cm^{-2}}$, respectively.
The common $3.9$--$5.0\,\mu$m wavelength window probes \bra, \brb, \paa, \pab, \ha, and \hb\  across the local Universe, cosmic noon, and reionization.
Our final sample contains 9146 galaxies, including 7883 in COSMOS and 1263 in GOODS-N.
 COSMOS constrains luminous emitters, while GOODS-N extends the luminosity functions toward fainter sources.
We measure dust-corrected line luminosities, account for line-detection completeness, and convert the six lines to equivalent \ha\ luminosities assuming Case B recombination.
Schechter fits reveal an increase in the characteristic \ha-equivalent luminosity toward cosmic noon. 
Integrating the resulting star formation rate functions above $0.24\,M_\odot\,\mathrm{yr}^{-1}$, we trace the cosmic star formation rate density from the local Universe to reionization.
The combined Brackett, Paschen, and Balmer measurements are broadly consistent with the canonical UV- and IR-based cosmic star formation history at $z\lesssim6$. Fixed-slope tests show that the H$\beta$ estimate at $z\sim8$ remains sensitive to the external photometric faint-end constraint and extrapolation below our spectroscopic limit.
These measurements provide complementary constraints on the distribution and integrated rate of star formation across cosmic time. 

\end{abstract}


\section{Introduction}
\label{sec:introduction}

Understanding the buildup of stellar mass across cosmic time remains a central goal of observational cosmology. 
Over the past two decades, extensive observational efforts have been devoted to reconstructing the cosmic star formation rate density (CSFRD), revealing a rise from the early Universe to a peak at $z \sim 2$--3, followed by a pronounced decline toward the present day \citep[e.g.,][]{Kennicutt_2012,Madau_2014}. 
This global star formation history provides a fundamental benchmark for models of galaxy formation and evolution, encapsulating the complex interplay between gas accretion, star formation, stellar feedback, and active galactic nucleus (AGN) activity.

A wide range of star formation rate (SFR) indicators have been developed to trace galaxy growth across cosmic time, including rest-frame ultraviolet (UV) continuum emission, infrared (IR) dust re-emission, and nebular recombination lines \citep[see reviews by][]{Kennicutt_2012, Madau_2014}. 
At high redshifts ($z \gtrsim 4$), constraints on the CSFRD have traditionally relied on rest-frame UV-selected Lyman-break galaxies \citep[e.g.,][]{Schenker_2013, Bouwens_2015, Finkelstein_2015}. 
While UV observations offer exceptional sensitivity to faint galaxies, their interpretation is complicated by dust attenuation and assumptions about star formation histories and the stellar initial mass function (IMF), which introduce systematic uncertainties in SFR estimates, particularly for massive and dusty systems.
Indeed, obscured star formation is known to contribute substantially to the CSFRD out to at least $z \sim 4$, and may remain important at even earlier epochs \citep{gruppioni_alpine-alma_2020, Zavala_2021, Algera_2023, fujimoto_alma_2024, Traina_2024,sun_2025,Lagache_2026,Traina_2026}.

Nebular recombination lines provide a powerful and complementary probe of star formation activity. 
Among them, the \ha\ emission line is widely regarded as one of the most robust tracers of ongoing star formation, as it directly traces ionizing photons from massive, short-lived ($<10$~Myr) stars and is less affected by dust extinction than rest-frame UV emission \citep[e.g.,][]{Kennicutt_2012, Tacchella_2022a}. 
Extensive narrow-band imaging and spectroscopic surveys have used \ha\ to measure luminosity functions (LFs) and CSFRDs at $z \lesssim 2.5$, establishing a well-calibrated reference at low and intermediate redshifts \citep[e.g.,][]{sobral_large_2013, Kashino_2013, Steidel_2014, Kriek_2015}. 

In recent years, \textit{James Webb Space Telescope} (JWST) has extended such studies to much higher redshifts.
With its unprecedented sensitivity and near- to mid-infrared wavelength coverage, JWST enables direct spectroscopic measurements of rest-frame optical emission lines from galaxies across much of cosmic history, notably with NIRSpec \citep[e.g.,][]{Shapley_2023_CEERS_Balmer,Curti_2024,Price_2024,deGraaff_2025_RUBIES,DEugenio_2025_JADES_DR3,Scholtz_2026_JADES_DR4}. 
Complementary JWST photometric studies constrain emission-line strengths and luminosity functions across a wide redshift range \citep[e.g.,][]{Williams_2023_JEMS,Rinaldi_2023,Endsley_2024,Shimakawa_2024_Pab_NB,Begley_2025_O3Hb_EW,NavarroCarrera_2026_Ha_photometry,Martis_2026_medband_HaLF}, including the lensed GLIMPSE measurements of \citet[hereafter K26]{Korber_2026_O3Hb_LF}.

In particular, the Near Infrared Camera (NIRCam) wide-field slitless spectroscopy (WFSS) mode \citep{Greene_2017, Rieke_2023} provides an efficient means of identifying emission-line galaxies without slit-target preselection over large contiguous redshift intervals \citep[e.g.,][]{Kashino_2023,Wang_ASPIRE_2023,Oesch_2023,Egami_2023,Backhaus_2024,Naidu_2024,Shen_2024_NEXUS,fu_medium-band_2025,Sun_2025_SAPPHIRES,Meyer_2026_COSMOS,Laishram_2026}. 
Free from slit losses and atmospheric absorption, slitless spectroscopy delivers emission-line-flux-limited samples of emission-line galaxies that are ideally suited for statistical measurements of LFs and cosmic star formation activity \citep[e.g.,][]{Sun_2022, Sun_2023, Matharu_2024}.
Using early NIRCam WFSS observations, \citet{Sun_2023} identified \ha\ emitters out to $z > 6$ and reported weak evolution in the \ha\ LF from $z \sim 2$ to 6. 
Subsequent Cycle-1 and Cycle-2 surveys, including FRESCO, CONGRESS, and the Medium-band Astrophysics with the Grism of NIRCam In Frontier Fields (MAGNIF) survey, have significantly expanded the census of \ha\ emitters at $3.8 < z < 6.6$, enabling direct measurements of the CSFRD beyond Cosmic Noon \citep{covelo-paz_h_2025, lin_luminosity_2025, fu_medium-band_2025}. 
These \ha-based CSFRD estimates are broadly consistent with UV-based measurements, with residual differences potentially reflecting selection effects, dust attenuation assumptions, and the different star-formation timescales probed by these tracers.
For the calibrations summarized in Table~1 of \citet{Kennicutt_2012}, \ha\ primarily traces stars younger than $\sim10$~Myr, whereas far-UV and near-UV emission remain sensitive to stellar populations with ages up to $\sim100$ and $\sim200$~Myr, respectively; IR emission can include contributions from stars with ages of $\sim100$~Myr or more, depending on the star-formation history and dust heating.

Despite these advances, existing emission-line-based measurements remain limited in two key aspects. 
First, most studies focus on a single emission-line tracer, typically \ha, which restricts the accessible redshift range within a given instrumental configuration. 
Second, CSFRD measurements across different redshifts are often derived using heterogeneous datasets, selection functions, and SFR calibrations, complicating efforts to construct a fully self-consistent picture of cosmic star formation over a wide redshift baseline.
In this work, we address these limitations by combining JWST/NIRCam F444W WFSS observations in COSMOS and GOODS-N. The large area of COSMOS-3D provides access to the relatively rare, luminous population, while the deeper GOODS-N observations extend our sensitivity to fainter emitters. The common F444W selection probes an observed wavelength window of approximately $3.9$--$5.0\,\mu\mathrm{m}$. 

From these multi-line samples, we measure LFs in successive redshift windows and derive the CSFRD on a common SFR scale, accounting for field-dependent selection functions and the line-specific external constraints described below.
Using multiple recombination lines extends the redshift coverage to $z\sim0$--9 and connects different cosmic epochs within a common observational configuration and SFR scale.
Our emission-line selection provides accurate spectroscopic redshifts. With spectroscopic redshifts, we can determine line luminosities and assign sources to redshift bins more reliably. We estimate the detection completeness through line-injection simulations. The resulting star formation rate functions (SFRFs) and CSFRD complement UV-, IR-, and SED-based studies on a common recombination-line scale.

This paper is organized as follows.
In \S\ref{sec:data}, we describe the JWST/NIRCam WFSS observations and the construction of the hydrogen recombination-line galaxy sample in the COSMOS and GOODS-N fields.
In \S\ref{sec:method}, we describe the emission-line flux measurements, SED modeling, calibration of the recombination lines as SFR tracers, and the completeness simulations used in our statistical analysis.
In \S\ref{sec:result}, we construct the luminosity functions and SFR functions for the recombination-line samples and derive the corresponding CSFRDs.
In \S\ref{sec:discuss}, we discuss the evolution of the star-forming galaxy population, compare our measurements with galaxy-formation models, and examine potential systematic uncertainties and environmental effects.
We summarize our main conclusions in \S\ref{sec:summary}.

Throughout this work, we assume a flat $\Lambda$CDM cosmology with $\Omega_{\mathrm{m}} = 0.3$, $\Omega_{\Lambda} = 0.7$, and $H_0 = 70\,\mathrm{km\,s^{-1}\,Mpc^{-1}}$, and adopt a Chabrier (2003) initial mass function.

\section{Data and Sample Construction} \label{sec:data}

We use emission-line catalogs from JWST/NIRCam WFSS observations in COSMOS and GOODS-N. From parent catalogs containing hydrogen, helium, and metal lines, we select H$\beta$, H$\alpha$, Pa$\beta$, Pa$\alpha$, Br$\beta$, and Br$\alpha$ emitters within the same line-dependent redshift intervals in both fields.
In this section, we summarize the data processing and sample construction, with further details provided in \citet{Fu_2026_COSMOS_ELG}.

\subsection{COSMOS-3D JWST/NIRCam WFSS Survey}
\label{data:c3d}

The main dataset used in this work comes from COSMOS-3D, a JWST Cycle 3 NIRCam imaging and wide-field slitless spectroscopy survey over a large contiguous survey area of $\sim0.33~{\rm deg}^2$ in the COSMOS field. COSMOS is one of the best-studied extragalactic fields, with extensive ancillary data from X-ray to radio wavelengths, deep optical and near-infrared imaging, and large spectroscopic redshift compilations. These data provide an essential foundation for identifying emission lines, validating redshifts, estimating photometric redshifts, and deriving galaxy physical properties.

\begin{figure}[!t]
    \centering
    \includegraphics[width=\columnwidth]{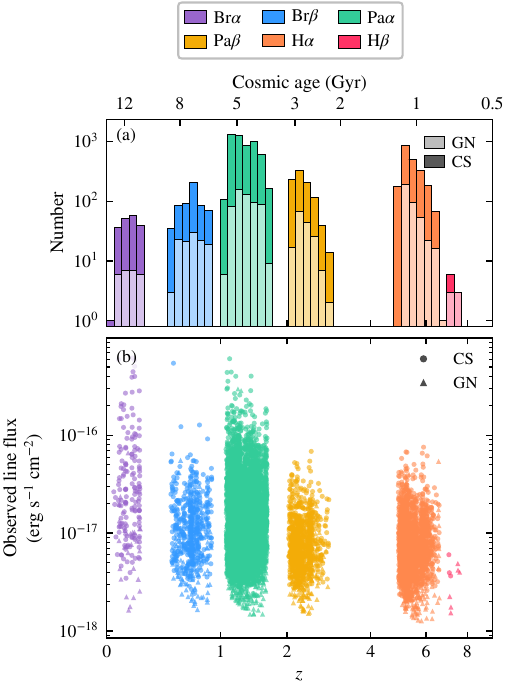}
    \caption{Redshift distribution and observed line fluxes of the six hydrogen recombination-line samples.
Upper panel: Unweighted source counts in bins of $\Delta\log_{10}(1+z)=0.02$, with stacked bars distinguishing COSMOS and GOODS-N.
Lower panel: Observed integrated line flux versus redshift for each detection, with circles for COSMOS and triangles for GOODS-N. Fluxes are shown without dust or Eddington-bias corrections.
Colors identify the emission lines in both panels.
}
    \label{fig:six_line_redshift_distribution}
\end{figure}

COSMOS-3D obtained NIRCam WFSS observations with the F444W filter. The F444W grism spectra cover approximately $3.9$--$5.0~\mu{\rm m}$, allowing the detection of different rest-frame optical and near-infrared emission lines across $z\sim0$--9. In the wavelength window adopted for this analysis, Br$\alpha$ traces galaxies at $z\lesssim0.3$ and Br$\beta$ at $z\simeq0.5-0.9$, Pa$\alpha$ and Pa$\beta$ trace galaxies around cosmic noon at $z\simeq1-3$, H$\alpha$ probes galaxies at $z\simeq5$--6, and H$\beta$ reaches galaxies at $z\simeq7$--9. The redshift distribution of the combined sample is shown in Figure~\ref{fig:six_line_redshift_distribution}. 

\subsection{GOODS-N: FRESCO and CONGRESS}
\label{data:goodsn}

We supplement COSMOS-3D with deeper NIRCam WFSS observations in GOODS-N. The F444W spectra are provided by the First Reionization Epoch Spectroscopically Complete Observations survey \citep[FRESCO;][]{Oesch_2023}, which has a grism footprint of $60.9~{\rm arcmin}^2$ in GOODS-N, with $7043$ s of grism integration per pointing. 

For source and emission-line identification in GOODS-N, we use both the FRESCO F444W spectra and the F356W spectra from the Complete NIRCam Grism Redshift Survey \citep[CONGRESS;][]{Egami_2023}. The additional wavelength coverage helps identify spectral features and establish redshifts. For the LF and CSFRD analysis, however, we include a source in a given hydrogen-line sample only when the corresponding line is detected in its F444W spectrum and lies within the same redshift interval adopted for COSMOS. The F356W spectra therefore support source identification but do not extend the redshift ranges of the six LF samples.
The typical $3\sigma$ integrated emission-line flux limits of the F444W spectra are approximately $3.2\times10^{-18}$ and $1.04\times10^{-18}~{\rm erg\,s^{-1}\,cm^{-2}}$ in COSMOS and GOODS-N, respectively.

For GOODS-N, we use the publicly released NIRCam mosaics and multi-band photometric catalog from Data Release 5 of the JWST Advanced Deep Extragalactic Survey \citep[JADES;][]{Eisenstein_2026_JADES_overview}, obtained from the DR5 release website.\footnote{\url{https://slate.ucsc.edu/~brant/jades-dr5/}} The NIRCam mosaics combine JADES observations with other public JWST programs and are described by \citet{Johnson_2026_JADES_DR5}; the source detection, photometric measurements, and uncertainty estimates are described by \citet{Robertson_2026_JADES_DR5}. We adopt the released images and catalog photometry rather than repeating the GOODS-N imaging reduction or photometric extraction. We use the same procedures for grism reduction, emission-line identification, line-flux measurement, and SED modeling in both fields.

\subsection{Imaging Data Reduction in the COSMOS Field}
\label{data:img_data_reduction}

In COSMOS, we use the available JWST/NIRCam imaging, including COSMOS-Web \citep{Cweb_survey_ref}, PRIMER, COSMOS-3D, and additional public programs and parallel observations where available. We also include HST/F814W imaging and deep ground-based optical imaging from HSC and the CFHT $U$ band. A detailed description of the COSMOS imaging reduction and photometric catalog construction is presented in \citet{Fu_2026_COSMOS_ELG}.

The NIRCam imaging data are reduced with the JWST Calibration Pipeline together with additional custom processing, including corrections for snowballs, wisps, $1/f$ noise, residual backgrounds, and astrometric offsets. The calibrated exposures are drizzled onto a common world coordinate system (WCS) grid with a pixel scale of $0\farcs03$. Empirical point-spread function (PSF) models are constructed for the JWST and HST images, and the images are PSF-matched to F444W.

We construct a $\chi^2$ detection image from the PSF-matched F115W, F150W, F200W, F277W, F356W, and F444W images, using the available bands at each position. Photometry is measured with SExtractor in dual-image mode using a cold+hot detection strategy. Fluxes are measured in Kron apertures with aperture corrections, and photometric uncertainties are estimated from random apertures in source-free regions. Ground-based HSC and CFHT photometry is measured using template-fitting deblending with the high-resolution F444W imaging as the positional and morphological prior.

The resulting multi-band photometric catalog is used for photometric redshift estimation, SED modeling, and emission-line identification. For WFSS extraction, we retain the COSMOS2025 \citep{COSMOS2025_cat} source IDs for matched objects and supplement them with bright sources detected in our catalog without COSMOS2025 counterparts, which are included to account for potential spectral contamination and reduce line misidentification \citep[see][]{Fu_2026_COSMOS_ELG}.

\subsection{NIRCam WFSS Data Reduction}
\label{data:wfss_data_reduction}
The spectroscopic data are based on the NIRCam WFSS observations obtained with the F444W filter and the Grism R in both fields. The GOODS-N F356W spectra additionally support source identification, as described in \S\ref{data:goodsn}. We reduce the WFSS exposures in both fields with a customized pipeline based on the NIRCam grism reduction procedures described by \citet{Sun_2023}, \citet{fu_medium-band_2025}, and \citet{Fu_2026_COSMOS_ELG}. The pipeline is designed to produce astrometrically corrected, background-subtracted, flux-calibrated 2D spectra for sources in the corresponding field's photometric catalog.

We first process the grism exposures with the JWST Stage-1 calibration pipeline. This step performs detector-level corrections and produces calibrated count-rate images. We then assign WCS information to the grism exposures and apply flat-field corrections using calibration flats matched to the same filter and detector configuration. For each filter, pupil, and module configuration, we construct a super-sky background frame from the flat-fielded grism images and subtract it from the individual exposures.

Because residual background structures remain after the super-sky subtraction, we perform an additional 2D sky-background subtraction. The background is estimated from source-masked and sigma-clipped grism images. To avoid biases from crowded regions or bright sources, we use a robust mode-like estimator when appropriate and switch to the median when the mode estimate becomes unreliable. We also remove residual $1/f$ noise by subtracting median background levels along both the row and column directions. This correction is applied iteratively on continuum-subtracted emission-line maps while masking high-significance pixels, so that real emission features are not over-subtracted.

Accurate astrometry is critical for NIRCam WFSS because small offsets can shift the spectral trace, wavelength solution, and flux calibration. We therefore refine the astrometry of the long-wavelength grism images using the simultaneously observed short-wavelength direct images. For COSMOS, bright and compact sources are detected in the short-wavelength images and matched to the COSMOS2025 catalog, which is registered to the HST astrometric frame. For GOODS-N, sources are matched to the JADES DR5 catalog. The measured offsets and rotations are then applied to the long-wavelength grism WCS. After the astrometric correction, the corresponding field's photometric catalog is projected onto each grism exposure to identify all sources that can produce spectra on the detector.

For each source and exposure, we determine the wavelength-dependent spectral trace using the calibrated tracing and dispersion solutions. We extract 2D spectra and calibrate the fluxes using the module- and grism-dependent sensitivity curves. Individual spectra from different exposures are resampled onto a common wavelength grid with $\Delta\lambda=1$ nm and coadded to form the final combined 2D spectrum for each source. The default extraction height in the spatial direction is 51 native long-wavelength pixels, corresponding to $3\farcs21$.

We also extract 1D spectra from the combined 2D spectra using a boxcar aperture with a height of 5 pixels, corresponding to $0\farcs31$. These 1D spectra are primarily used for emission-line detection and redshift identification. The aperture is not intended to capture the total flux of extended sources, but it provides spectra with a high signal-to-noise ratio (SNR) for detecting emission features. Final line fluxes are instead measured from the 2D spectra, which better preserve the spatial distribution of the line emission and account for morphology broadening in slitless spectroscopy.

\subsection{Sample Construction}
\label{data:sample_construction}

The parent emission-line sample is constructed through a multi-step procedure that includes line detection, redshift estimation, line identification, visual inspection, and line-flux measurement. 
Emission-line candidates are first detected in both 1D and 2D continuum-subtracted spectra. The continuum is removed with a median filter, and the residual spectra are smoothed with a Gaussian kernel matched to the instrumental resolution. Candidate lines are required to be detected in the 1D spectra with ${\rm SNR}>3$ and to have consistent 2D emission features near the expected spectral trace. A separate 1D detection procedure is also used to recover broader features and extended low-redshift emission lines that may have low per-pixel signal-to-noise ratios in the 2D spectra.

A major challenge in slitless spectroscopy is source confusion. Since spectra from nearby sources can overlap, a detected emission feature may be spatially consistent with multiple possible emitters. The parent catalog therefore identifies each line using information from the grism spectra, direct images, multi-band SEDs, photometric redshift probability distributions, observed equivalent widths, external spectroscopic redshifts, and the spatial consistency between the direct image and the 2D line profile. This identification step combines automatic template matching, visual inspection, a Siamese neural network for morphology matching, and a large language model (LLM)-assisted classification framework. A detailed description of the source-association procedure and quantitative validation is presented in \citet{Fu_2026_COSMOS_ELG}. The Siamese network maps paired direct-image and 2D emission-line cutouts into a common feature space using convolutional neural networks, where smaller separations indicate greater morphological similarity and help distinguish the true emitter from nearby contaminants. The LLM combines these morphological scores with spectral-template redshifts, photometric constraints, and emission-line equivalent widths to propose line identifications and assign likely emitting sources. The final line identifications are visually inspected.


\begin{figure*}[!t]
    \centering
    \includegraphics[width=\textwidth]{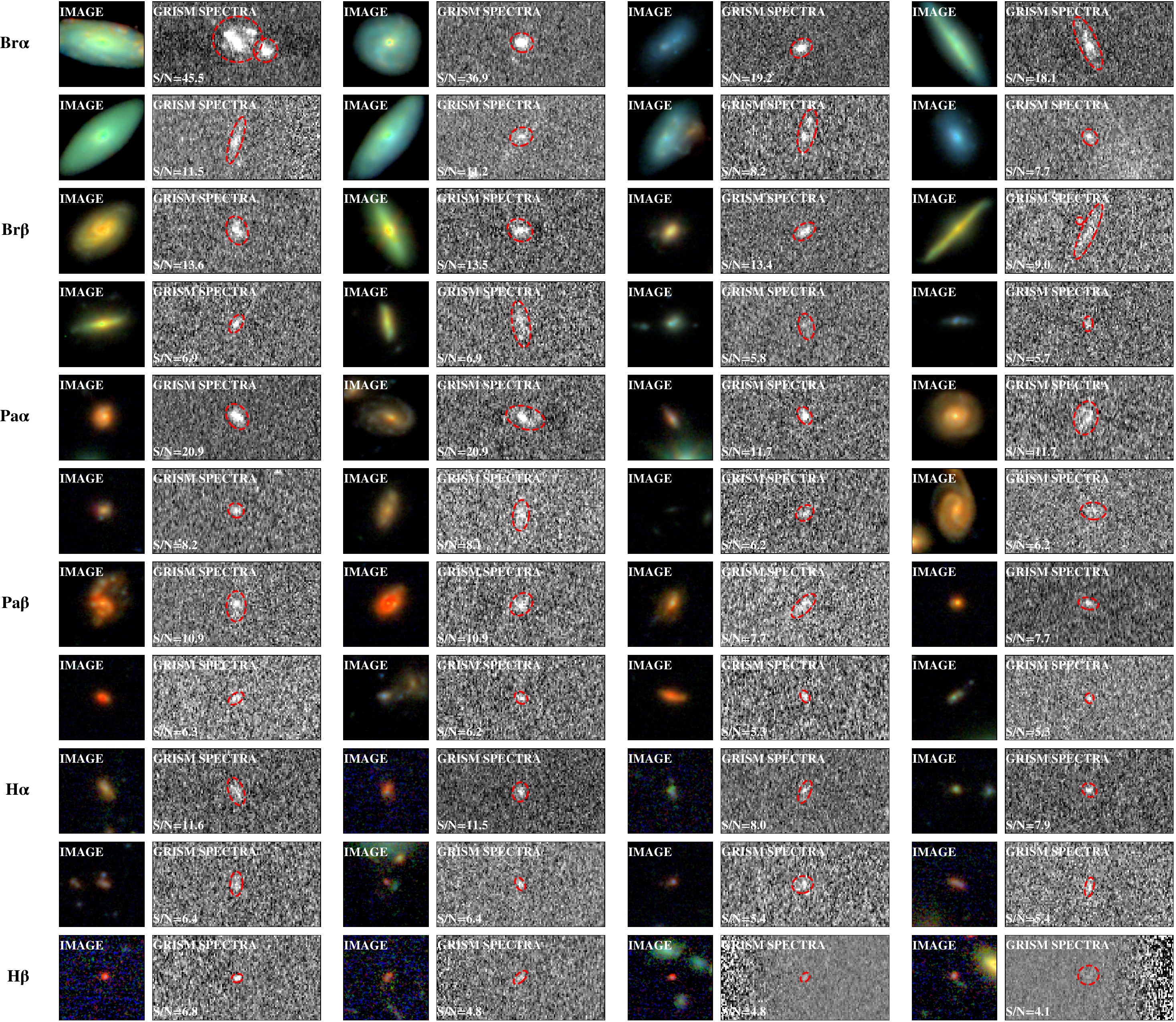}
    \caption{Examples of hydrogen recombination-line galaxies selected from the COSMOS field (COSMOS-3D).
Each row shows one line: Br$\alpha$, Br$\beta$, Pa$\alpha$, Pa$\beta$, H$\alpha$, or H$\beta$. For each line, eight sources (four for H$\beta$) are shown in decreasing line SNR. Each source has a multi-band RGB image and the corresponding 2D grism spectrum. The RGB channels use F115W+F150W (blue), F200W+F277W (green), and F356W+F444W (red). Images are rotated to align with the dispersion direction. The line SNR is labeled in each spectrum, and red ellipses mark the line-flux measurement apertures.
    }
    \label{fig:example_cutout}
\end{figure*}

We select hydrogen recombination-line emitters from the parent catalog including Br$\alpha$, Br$\beta$, Pa$\alpha$, Pa$\beta$, H$\alpha$, and H$\beta$. 
Examples of randomly selected COSMOS sources spanning a range of line signal-to-noise ratios are shown in Figure~\ref{fig:example_cutout}. 

The final LF sample contains 9146 galaxies: 7883 in COSMOS and 1263 in GOODS-N. It includes 184 Br$\alpha$ emitters, 574 Br$\beta$ emitters, 5324 Pa$\alpha$ emitters, 942 Pa$\beta$ emitters, 2113 H$\alpha$ emitters, and 9 H$\beta$ emitters. GOODS-N contributes 26, 118, 567, 165, 381, and 6 sources to these six samples, respectively. These counts refer to the source samples entering the LF analysis after the uniform effective-completeness requirement $C_{\rm eff}>0.01$ (\S\ref{res:line_lfs}). We exclude AGNs whose recombination-line profiles are dominated by broad emission, because the weaker narrow-line component cannot be reliably separated in our slitless spectra to constrain star formation in their hosts. We fit single- and double-Gaussian models to each continuum-subtracted 1D line profile, select the preferred model using the Bayesian information criterion, and flag broad features if any component has a full width at half maximum (FWHM) greater than $1000~\mathrm{km\,s^{-1}}$. We use compactness ($F(0\farcs7)/F(0\farcs3)<2$) and visual inspection to distinguish AGN candidates from sources whose spatial extent along the dispersion direction broadens their emission lines.
The final hydrogen-line sample provides the basis for constructing line LFs, converting them to SFRFs, and tracing the CSFRD across $z\sim0$--9.

\section{Method} \label{sec:method}

In this section, we describe the methodology used to characterize the physical properties and star formation activity of our emission-line galaxy sample. 
We first measure emission-line fluxes from the grism spectra (\S\ref{method:flux}). 
We then perform SED modeling using the multi-band photometry to infer galaxy properties (\S\ref{method: SED}). 
Next, we compare the measured recombination-line luminosities with theoretical values, accounting for dust attenuation (\S\ref{method:sfr_tracers}). 
Finally, we quantify the line-detection completeness through injection--recovery simulations and derive completeness corrections that are incorporated into the subsequent statistical analyses (\S\ref{method:Completeness}).

\subsection{Line Flux Measurement} \label{method:flux}  

We use \textsc{photutils} \citep{photutils} on the 2D spectra to perform source detection and measure the fluxes of emission lines.
While the median-filter--based continuum subtraction is sufficient for emission-line detection, it can lead to flux loss due to oversubtraction in the vicinity of the line.
To mitigate this effect, we adopt a more precise procedure to model and subtract the local continuum prior to flux measurements.

We fit a cubic polynomial to the continuum in each spatial row of the 2D spectra, using regions adjacent to the emission line ($\pm 60$ pixels, corresponding to $\pm 600$~\AA).
Source detection is subsequently performed on the continuum-subtracted spectra to generate a segmentation map, which is used to mask pixels associated with the emission line.
The continuum fitting is then repeated with the masked regions excluded.
This procedure is iterated until convergence, after which the emission line is measured from the final continuum-subtracted spectra.
To account for both the spatial distribution and the morphology broadening effect of the slitless spectra, we obtain Kron-aperture photometry \citep{Kron_1980} with a Kron factor of 3, which is large enough to include nearly all flux while maintaining a high SNR. Examples of the Kron apertures are shown as red ellipses in the 2D spectra in Figure~\ref{fig:example_cutout}. 

Note that emission in the outer regions of a line map may fall below the detection SNR threshold but still contribute to the total SFR. On the other hand, noise can preferentially scatter faint emission lines above the detection threshold, leading to overestimated fluxes and potentially biasing the inferred LF \citep{eddington1913}. We calibrate a flux-correction relation separately for each field and line using the injection--recovery simulations described in \S{\ref{method:Completeness}}. In each bin of input log flux, we calculate the median recovered log flux for the detected injections. Within the calibrated range, we then linearly interpolate the tabulated input log flux as a function of this median recovered log flux, evaluating the relation at the measured flux of each galaxy. We apply this relation to the measured line fluxes as a statistical correction for median recovery biases.  

\subsection{SED modeling} \label{method: SED}

We perform SED modeling with \textsc{bagpipes} \citep{Carnall_2018} for the emission-line galaxies. 
For each source, the redshift is fixed to the spectroscopic redshift. 
For COSMOS, we use the available multi-band photometry from Subaru/HSC, HST/ACS, and JWST/NIRCam described in \S\ref{data:img_data_reduction}; for GOODS-N, we use the HST and JWST photometry from JADES DR5. The same SED-model assumptions are applied to both fields. 
We also include the observed hydrogen recombination-line fluxes in the fit to better constrain the recent star formation activity. 
We use the \citet{Bruzual_2003} stellar population synthesis models, and adopt the non-parametric continuity star formation history implemented in \textsc{bagpipes}. 
The star formation history is divided into age bins with edges at 0, 3, 10, and 100 Myr, followed by three broader bins between 100 Myr and the age of the Universe at the source redshift \citep{Leja_2019}. 
The ratios of the SFRs between adjacent time bins are allowed to vary over a broad range, with a Student-$t$ prior imposed on each \texttt{dsfr} parameter. 
The stellar metallicity is allowed to vary from 0.001 to 2.5 $Z_\odot$, with a logarithmic prior.
Dust attenuation is modeled using the Salim attenuation law \citep{Salim_2018}. 
We allow the $V$-band attenuation to vary over $A_V=0$--4 mag. 
The extra attenuation in birth clouds is controlled by the parameter $\eta$, which is allowed to vary between 1 and 3. 
We also allow the slope modification of the attenuation curve to vary over $\delta=-0.6$--0.2, and the strength of the UV bump to vary over $B=0$--5. 
Nebular emission is included, with the ionization parameter allowed to vary over $\log U=-4$--0.

For each fitted source, we derive posterior distributions for the model parameters and physical quantities, including stellar mass, 10 Myr-averaged SFR, metallicity, dust attenuation, and ionization parameter. For the dust correction of emission lines, we also derive posterior distributions of $A_{\rm line}$ for H$\alpha$, H$\beta$, Pa$\alpha$, Pa$\beta$ and Br$\beta$, and obtain their median values and $1\sigma$ uncertainties. We apply these attenuation corrections to the line-based SFRs and propagate the $A_{\rm line}$ uncertainties into the SFR errors. Dust attenuation at the wavelength of Br$\alpha$ is negligible, we therefore apply no dust-attenuation correction to this line.

\subsection{Star Formation Rate Tracers}
\label{method:sfr_tracers}

We convert the dust-corrected luminosity of each hydrogen line to an H$\alpha$-equivalent luminosity using its intrinsic Case B ratio. This assumes that, for each line, the dust-corrected ratio $L_{\rm line}/L_{\rm H\alpha}$ is approximately consistent with the corresponding Case B value across our sample. Comparing Balmer, Paschen, and Brackett lines tests this assumption across a wide wavelength range, since dust attenuates the Balmer lines more strongly \citep[e.g.,][]{Gimenez_2022,Wozniak_2026,Baron_2026}.

We construct a calibration sample with simultaneous measurements of H$\alpha$, H$\beta$, and at least one Paschen or Brackett line. For the Paschen-line sample, we search the NIRSpec spectra available in the Dawn JWST Archive (DJA; \citealp{Heintz_2024_DJA,deGraaff_2025_RUBIES})\footnote{\url{https://dawn-cph.github.io/dja/}} and select sources with detections of H$\alpha$, H$\beta$, and either Pa$\alpha$ or Pa$\beta$. The archival NIRSpec spectra were processed with \textsc{msaexp} \citep{Brammer_2022_msaexp}. The Pa$\alpha$ sample covers the redshift range $z\simeq0.4$--2, while the Pa$\beta$ sample covers $z\simeq0.5$--3. For the Brackett-line sample, we start from the Br$\alpha$ and Br$\beta$ emitters in our COSMOS NIRCam grism sample and cross-match them with public optical spectroscopic redshifts and spectra from Data Release 1 of the Dark Energy Spectroscopic Instrument (DESI DR1; \citealp{DESI_2026_DR1}) and zCOSMOS \citep{Lilly_2007}. This allows us to obtain the corresponding H$\alpha$ and H$\beta$ measurements for the same galaxies.

We fit Gaussian profiles to H$\alpha$, H$\beta$, and the Paschen lines, estimating the local continuum from adjacent line-free windows and propagating the spectral errors through the fits. We mask contaminated pixels, inspect the fits visually, and exclude sources with unreliable fits, severe contamination, or $\mathrm{SNR}<3$ in the relevant lines.
We use the BPT diagram, based on [O\,III]$\lambda5007$/H$\beta$ and [N\,II]$\lambda6584$/H$\alpha$, to exclude AGN candidates lying above the theoretical maximum-starburst boundary of \citet{Kewley_2001}.
For sources with Brackett-line measurements from our NIRCam grism spectra and H$\alpha$ and H$\beta$ measurements from DESI or zCOSMOS spectra, aperture losses in the ground-based spectra should also be taken into account.
For the comparison with DESI/zCOSMOS H$\alpha$ fluxes, we measure the grism Brackett fluxes in an aperture-matched manner. We restrict the grism measurement to the sky region corresponding to the DESI fiber or zCOSMOS slit footprint. 

We estimate the nebular dust attenuation from the Balmer decrement. We adopt the \citet{Salim_2018} attenuation curve with the slope-modification parameter fixed at $\delta=-0.45$, corresponding to an approximately SMC-like attenuation curve. Assuming an intrinsic H$\alpha$/H$\beta$ ratio of 2.86, we derive the color excess of the nebular gas as
\begin{equation}
E(B-V)_{\rm neb} =
\frac{2.5}{k({\rm H}\beta)-k({\rm H}\alpha)}
\log_{10}
\left[
\frac{({\rm H}\alpha/{\rm H}\beta)_{\rm obs}}
{({\rm H}\alpha/{\rm H}\beta)_{\rm int}}
\right] .
\end{equation}
where $k(\lambda)$ is the attenuation curve evaluated at the wavelength of each line. We then correct the observed luminosity of each recombination line using
\begin{equation}
L_{\rm corr}(\lambda)
=
L_{\rm obs}(\lambda)
\,10^{0.4\,E(B-V)_{\rm neb}\,k(\lambda)} .
\end{equation}

\begin{figure*}[!t]
    \centering
    \includegraphics[width=\textwidth]{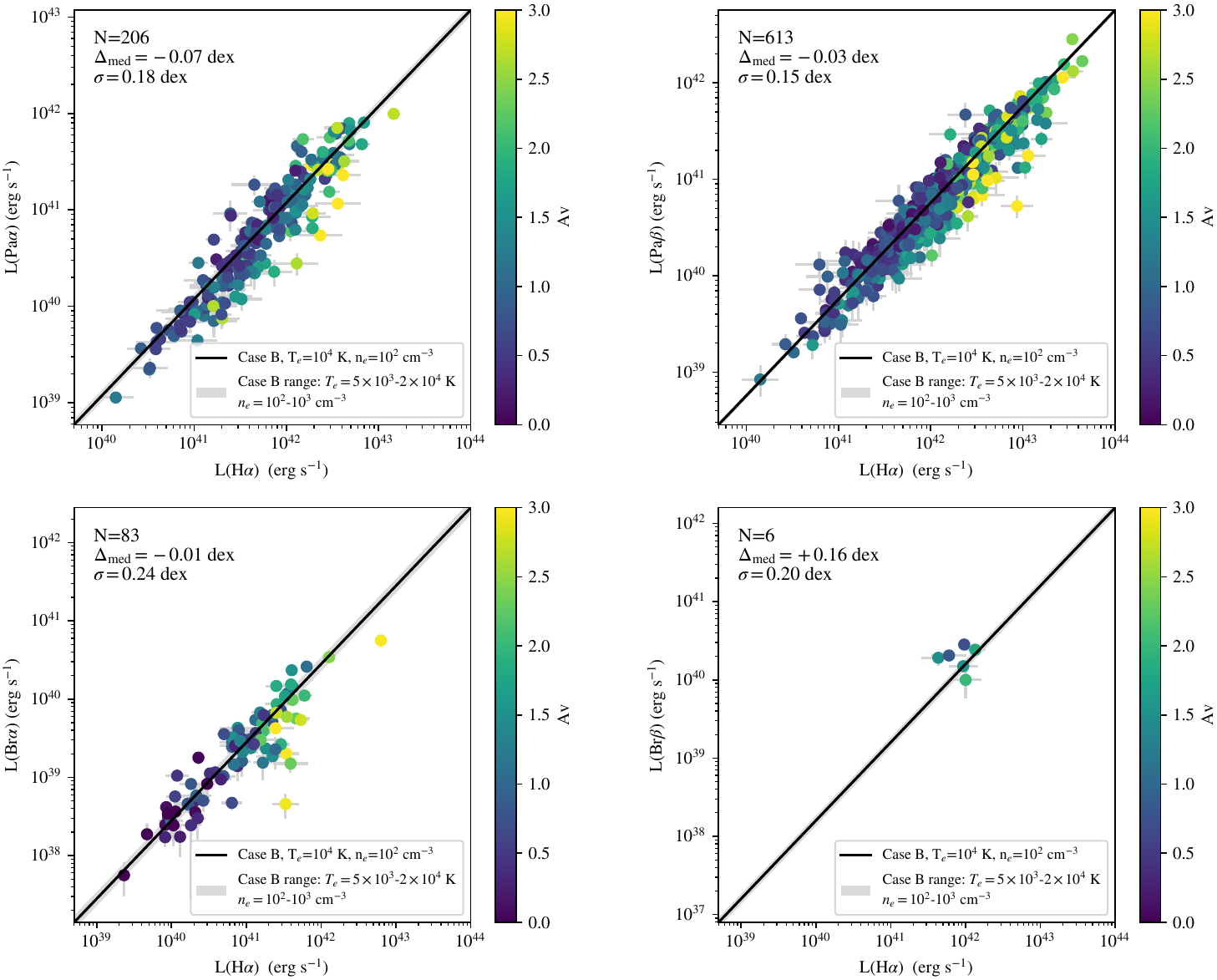}
    \caption{Comparison of dust-corrected recombination-line luminosities for the COSMOS-field calibration sample.
The panels show Pa$\alpha$, Pa$\beta$, Br$\alpha$, and Br$\beta$ luminosities versus H$\alpha$ luminosity. Points are color-coded by the Balmer-decrement-based $A_V$, and gray error bars show the luminosity uncertainties. Black solid lines indicate the fiducial Case B ratios for $T_e=10^4~\mathrm{K}$ and $n_e=10^2~\mathrm{cm^{-3}}$. Gray bands show the minimum--maximum theoretical ratios over $T_e=5\times10^3$--$2\times10^4~\mathrm{K}$ and $n_e=10^2$--$10^3~\mathrm{cm^{-3}}$, calculated with \textsc{PyNeb} using the recombination emissivities of \citet{Storey_1995}. In each panel, $N$ gives the number of sources, $\Delta_{\rm med}$ is the median logarithmic offset from the fiducial relation, and $\sigma$ is half the 16th--84th percentile width of these offsets. }
    \label{fig:line_luminosity_comparison}
\end{figure*}

Figure~\ref{fig:line_luminosity_comparison} compares the dust-corrected luminosities of Pa$\alpha$, Pa$\beta$, Br$\alpha$, and Br$\beta$ with the dust-corrected H$\alpha$ luminosity.  
We calculate the Case B recombination line ratios over $T_e=5\times10^3$--$2\times10^4~\mathrm{K}$ and $n_e=10^2$--$10^3~\mathrm{cm^{-3}}$ with \textsc{PyNeb} \citep{Luridiana_2015}. The ratios use the hydrogen recombination emissivities tabulated by \citet{Storey_1995}. The black solid lines in each panel show the fiducial ratios for $T_e=10^4~\mathrm{K}$ and $n_e=10^2~\mathrm{cm^{-3}}$ \citep[e.g.,][]{Osterbrock_2006}, as summarized in Table~\ref{tab:line_ratio}, while the gray bands span the minimum and maximum ratios over the adopted temperature and density ranges. The narrow bands indicate that variations in electron temperature and density over these representative H~{\sc ii}-region conditions produce only modest changes in the theoretical line ratios.
The dust-corrected luminosities broadly follow the fiducial Case B relations, with object-to-object scatter beyond the theoretical bands. Each panel reports the median offset and scatter.
This comparison supports the adopted Case B ratios and Balmer-decrement corrections for the calibration sample, although galaxy-to-galaxy variations in the attenuation curve, such as its slope-modification parameter $\delta$, may contribute to the scatter. We use the fiducial ratios to convert our line luminosities to H$\alpha$-equivalent luminosities and derive SFRs with the calibration in \S\ref{res:line_sfrfs}. 

\begin{table}[t]
\centering
\caption{Hydrogen Recombination-line Ratios}
\label{tab:line_ratio}
\begin{tabular*}{0.48\textwidth}{@{\hspace{1cm}}c@{\hspace{3cm}}c}
\hline\hline
Emission Line & $L_{\rm line}/L_{\mathrm{H}\alpha}$ \\
\hline
H$\beta$   & 0.3493 \\
Pa$\alpha$ & 0.1183 \\
Pa$\beta$  & 0.0570 \\
Br$\alpha$ & 0.0280 \\
Br$\beta$  & 0.0159 \\
\hline
\end{tabular*}

\vspace{2pt}
\begin{minipage}{\columnwidth}
\footnotesize
\textit{Note.} Intrinsic hydrogen recombination-line luminosity ratios
relative to H$\alpha$, calculated with \textsc{PyNeb} assuming Case B
recombination, $T_e=10^4$ K, and $n_e=100~{\rm cm^{-3}}$.
\end{minipage}
\end{table}

\begin{figure}[!t]
    \centering
    \includegraphics[width=\columnwidth]{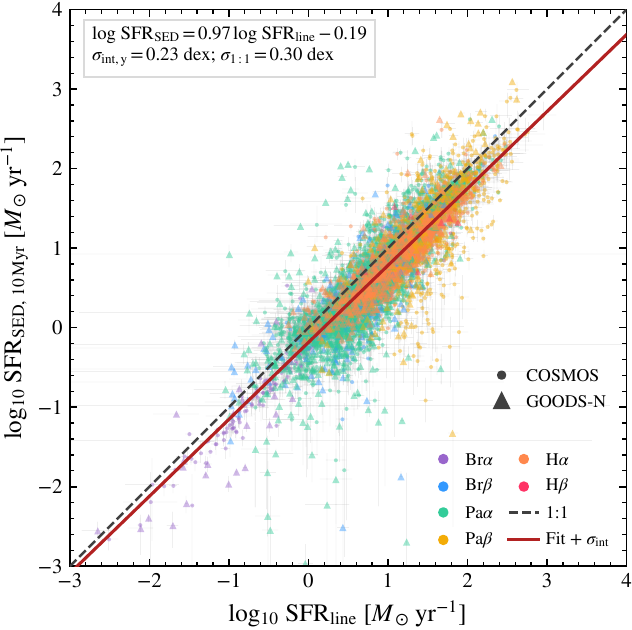}
    \caption{Comparison of recombination-line and 10 Myr-averaged SED-based SFRs for COSMOS (dots) and GOODS-N (triangles). Colors identify the six hydrogen recombination lines. Gray error bars include dust-corrected line-flux uncertainties on the horizontal axis and SED posterior uncertainties on the vertical axis. The dashed line marks equality; the red solid line is the best-fitting relation for the combined data, with a free intrinsic vertical scatter $\sigma_{\rm int,y}$. The annotations also give the observed standard deviation of $\log_{10}({\rm SFR}_{\rm SED}/{\rm SFR}_{\rm line})$, denoted by $\sigma_{1:1}$, without subtracting measurement errors.}
    \label{fig:SFR_line_vs_SFR_SED}
\end{figure}

We also compare recombination-line SFRs with the SED-based SFRs averaged over the most recent 10 Myr (Figure~\ref{fig:SFR_line_vs_SFR_SED}). For this comparison, the line luminosities are corrected for flux measurement biases and dust attenuation before conversion to SFR using the calibration in \S\ref{res:line_sfrfs}. 
We fit $y=ax+b$, where $x=\log_{10}{\rm SFR}_{\rm line}$ and $y=\log_{10}{\rm SFR}_{\rm SED}$, with a Gaussian likelihood that includes measurement errors in both axes and a free intrinsic scatter in $y$. 
Fitting all line emitters from both fields jointly with one shared intrinsic-scatter parameter gives $(a,b)=(0.968,-0.185)$ and $\sigma_{\rm int,y}=0.228$ dex. Thus, the two SFR estimates are nearly proportional. Because the SED modeling includes the measured line fluxes and supplies the dust corrections, the two estimates are not independent. This comparison therefore tests internal consistency rather than providing an independent SFR calibration. 

\subsection{Completeness correction} \label{method:Completeness}
To determine the LFs, it is essential to consider the completeness of our source detection. 
We first construct a grid of artificial source positions across the field, with a spacing of 10 arcsec in both the R.A. and Dec. directions. 
At each grid position, we extract the local 2D spectra from the individual grism exposures using the same trace model, wavelength solution, bad-pixel mask, and inverse-variance maps as used for the real sources. 
This procedure preserves the spatially varying depth, contamination residuals, detector defects, and wavelength coverage of the real data.

We then inject mock emission-line sources into the extracted 2D spectra of each individual grism exposure, before resampling and coadding them in the same way as the real data. 
For each line species, we insert 20 mock lines at each grid position in the fiducial simulation with a distribution of input line flux, equivalent width and continuum slope, which are randomly drawn from the empirical distributions measured from our observed line sample. 
The line center is drawn uniformly over the wavelength coverage of F444W filter. 
The continuum is modeled as a power law, $f_{\lambda}\propto\lambda^{\alpha}$, normalized to reproduce the continuum magnitude implied by the input line flux and equivalent width. 
For H$\alpha$ and H$\beta$, the spatial profile of the injected line is modeled as an elliptical Gaussian. 
Because Paschen and Brackett lines correspond to relatively low-redshift emitters in our wavelength range, these sources can be more spatially extended and morphologically complex than compact high-redshift Balmer-line emitters. 
For Pa$\alpha$, Pa$\beta$, Br$\alpha$, and Br$\beta$, we therefore use real F444W image cutouts as spatial templates. 
In this case, the segmentation map is used to isolate the target source, and the image cutout is resampled to the grism pixel scale and normalized before being scaled to the desired input line flux. 
The use of real image cutouts therefore captures the realistic source sizes, asymmetries, and surface-brightness distributions of lower-redshift line emitters, all of which can influence line recovery in slitless spectra.

We add the corresponding Poisson noise from the injected source and update the inverse-variance map of each exposure accordingly. 
The simulated spectra from the individual exposures are then coadded in the same way as the real data.
We then run the same 1D and 2D emission-line detection algorithms as used for the science catalog.   
A mock line is considered recovered only if both the 1D and 2D detections are present, and the recovered line center is within two instrumental FWHM of the input wavelength. 
Because the mock lines are inserted into real WFSS spectra, some positions may contain pre-existing residual line-like features from contamination or imperfect background subtraction. 
We therefore run the line finder on the original spectra before injection and record these features. 
A recovered feature is counted as a successful recovery only if it is matched to the injected line rather than to a pre-existing feature in the original spectrum.

\begin{figure*}[!t]
    \centering
    \includegraphics[width=\textwidth]{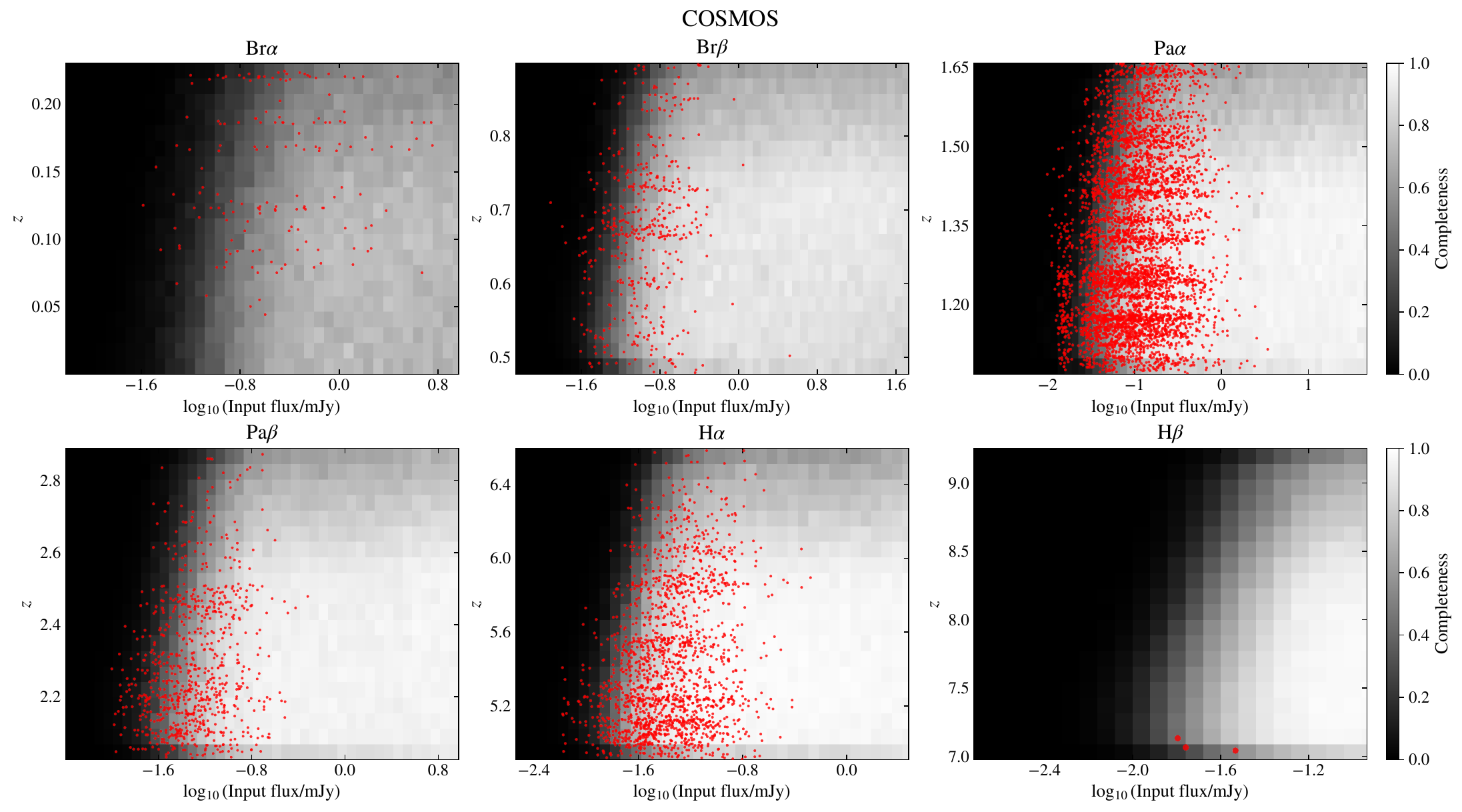}\par
    \vspace{4pt}
    \includegraphics[width=\textwidth]{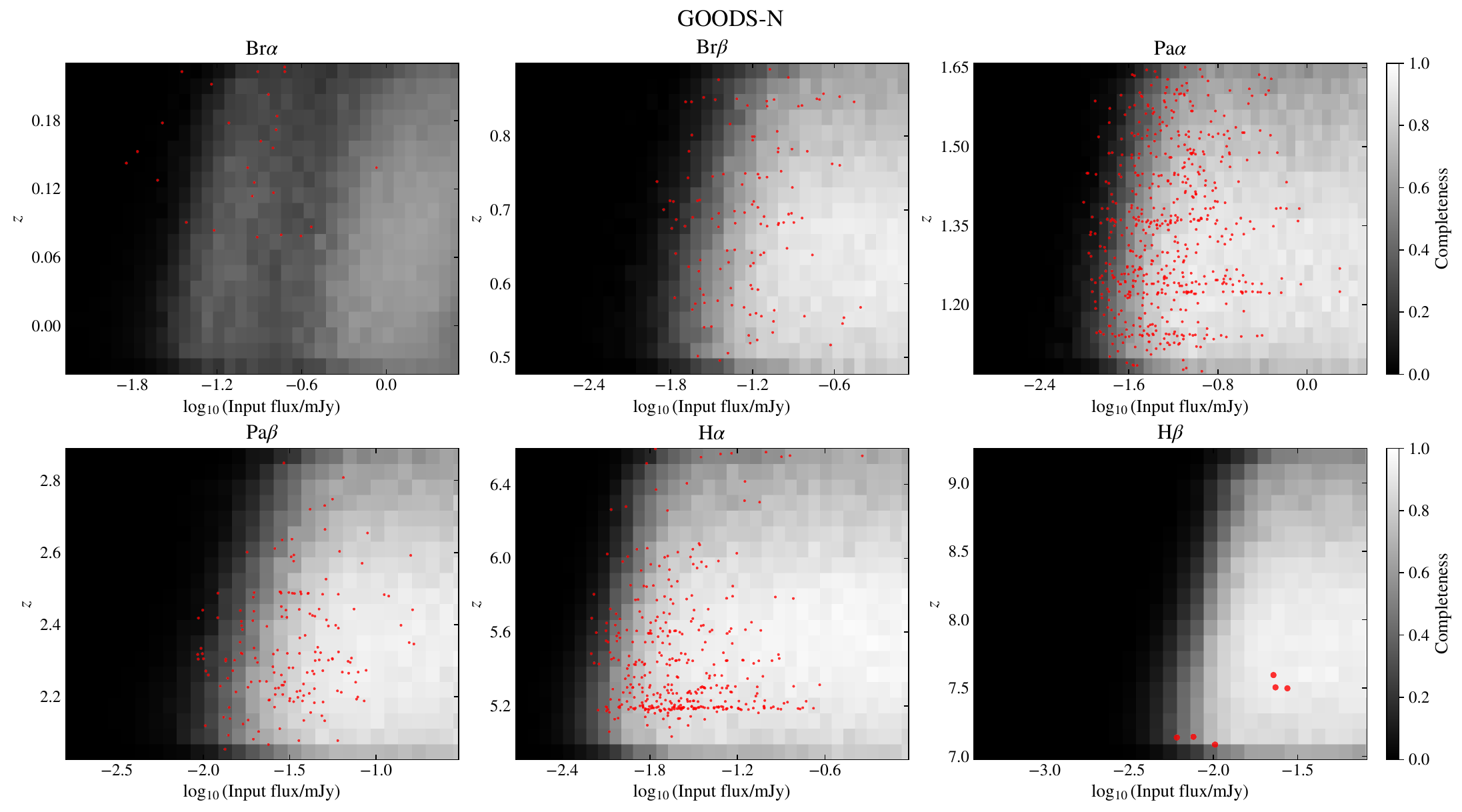}
    \caption{Emission-line detection completeness in COSMOS (upper) and GOODS-N (lower), as a function of redshift and input line flux, derived from the injection--recovery simulations. For each field we show the six hydrogen recombination lines on the same 0--1 completeness scale.
Red points mark the redshifts and flux-deboosted, but not dust-corrected, fluxes of sources in the corresponding field. The LF calculation uses the field-specific completeness grids and combines their effective volumes.
    }
    \label{fig:completeness}
\end{figure*}

For each field and line, we estimate completeness in input-flux and redshift cells as $C_f(F,z)=N_{\rm rec}/N_{\rm inj}$, where recovered injections satisfy the adopted detection and quality criteria. Figure~\ref{fig:completeness} shows the separate completeness grids for COSMOS and GOODS-N.
For the low-redshift Brackett sources, the completeness does not always reach unity even at moderately high total line fluxes. This is mainly due to the complex morphologies of low-redshift galaxies and the spatially uneven distribution of their star-forming regions. In some cases, the Brackett-line emission is offset from the center of the continuum image, for example when the dominant star-forming region lies in the outer disk. Since the detection is performed on the grism spectra extracted around the continuum source position, such spatial offsets can reduce the recovered line signal and may cause the emission line to be missed.

\section{Results} \label{sec:result}

\subsection{Line luminosity functions}
\label{res:line_lfs}

We construct dust-corrected LFs for H$\beta$, H$\alpha$, Pa$\beta$, Pa$\alpha$, Br$\beta$, and Br$\alpha$ using the combined COSMOS+GOODS-N samples. Both fields use the same line-dependent redshift intervals, corresponding to the adopted F444W window $3.879$--$4.986\,\mu$m.

For a source with flux-deboosted, but not dust-corrected, luminosity $L_{\rm att}$, the effective volume in field $f$ is
\begin{equation}
V_{{\rm eff},f}(L_{\rm att})=\int_{z_{\min}}^{z_{\max}}
C_f[F(L_{\rm att},z),z]\,\frac{dV_f}{dz}\,dz .
\end{equation}
The completeness grids incorporate the field-dependent coverage and recovery probability. The total accessible volume is $V_{\rm eff}=V_{{\rm eff},{\rm CS}}+V_{{\rm eff},{\rm GN}}$, including the contribution from a field even when that field contains no detected source at the luminosity in question. To avoid extreme inverse-volume weights, we uniformly require
\begin{equation}
C_{\rm eff}(L_{\rm att})\equiv
\frac{V_{{\rm eff},{\rm CS}}+V_{{\rm eff},{\rm GN}}}
{V_{{\rm geom},{\rm CS}}+V_{{\rm geom},{\rm GN}}}>0.01
\label{eq:ceff_cut}
\end{equation}
for all six lines. This is a volume-averaged joint completeness, not the local recovery fraction evaluated at the source redshift.

We bin the dust-corrected luminosities in intervals of $\Delta\log_{10}L=0.3$ dex and estimate
\begin{equation}
\Phi_k=\frac{1}{\Delta\log_{10}L_k}\sum_{i\in k}
\frac{1}{V_{{\rm eff},{\rm CS}}(L_{{\rm att},i})+V_{{\rm eff},{\rm GN}}(L_{{\rm att},i})}.
\end{equation}
Dust correction changes the luminosity used for binning, but not the flux used to evaluate the selection function. Table~\ref{tab:dustcorr_LF} reports the binned LFs.

The covariance includes weighted Poisson noise, field-dependent cosmic variance, and the finite number of injection--recovery trials. We propagate the last term with 1000 realizations of the completeness grids, drawing calibrated cells from $\mathrm{Beta}(N_{\rm rec}+1/2,N_{\rm inj}-N_{\rm rec}+1/2)$. The source list, bin boundaries, and nominal completeness selection remain fixed in these realizations. A shared grid realization is used for all sources in a field, preserving correlations between LF bins. Cosmic variance is correlated between bins within a field, while COSMOS and GOODS-N are treated as independent sightlines.
We estimate cosmic variance by integrating the matter power spectrum over a rectangular survey window set by each field's area and line-dependent redshift interval \citep[cf.][]{Moster_2011}, assuming a 2:1 angular aspect ratio and the \citet{Tinker_2010_bias} bias. The estimated fractional fluctuations span $8$--$21\%$ in COSMOS and $15$--$40\%$ in GOODS-N. We combine the independent fields through their luminosity-dependent effective-volume responses and include the resulting cosmic-variance covariance in the LF fits. The combined fractional cosmic-variance uncertainties span approximately $7$--$23\%$ across the populated luminosity bins of the six line samples.

We fit the LFs with a Schechter function per dex,
\begin{equation}
\Phi(L)=\ln 10\,\phi^\ast
\left(\frac{L}{L^\ast}\right)^{\alpha+1}
\exp\left(-\frac{L}{L^\ast}\right),
\end{equation}
where $\phi^\ast$, $L^\ast$, and $\alpha$ are the normalization, characteristic luminosity, and faint-end slope. All fits are performed on an H$\alpha$-equivalent luminosity scale using the Case B ratios in Table~\ref{tab:line_ratio}. We fit the binned LFs using a Gaussian likelihood in $\log_{10}\Phi$:
\begin{equation}
\ln\mathcal{L}=-\frac{1}{2}\left[
\boldsymbol{r}^{\mathsf T}\boldsymbol{\Sigma}^{-1}\boldsymbol{r}
+\ln|\boldsymbol{\Sigma}|+n\ln(2\pi)\right],
\end{equation}
where the residuals compare the measured log LF with the model at the bin centers, $\boldsymbol{\Sigma}$ includes the full covariance of our bins, and $n$ is the number of LF bins included in the fit.

We adopt the following choices of external information.
\begin{enumerate}
\item \textit{H$\alpha$, Pa$\alpha$, Pa$\beta$, and Br$\beta$: our data only.} We fit the combined COSMOS+GOODS-N bins without adding any literature LF points or informative priors. All three Schechter parameters are free. The wide COSMOS area and deeper GOODS-N data supply complementary luminosity coverage.
\item \textit{H$\beta$: our data plus K26.} Our H$\beta$ sources populate only four nonempty bins, providing limited leverage on the faint-end slope. 
Few spectroscopic LF measurements are available at these redshifts.
We therefore include the eight $7<z<8$ bins ($\langle z\rangle=7.48$) of the dust-corrected H$\beta$ LF in Table~C.5 of K26. Their independent, lensed GLIMPSE field extends to substantially lower luminosities. These are photometrically inferred measurements of the blended H$\beta$+[O\,III] emission, separated using an empirical $R_3(M_{\rm UV})$ relation, where $R_3\equiv L([\mathrm{O\,III}]\lambda5008)/L(\mathrm{H}\beta)$ and $M_{\rm UV}$ is the rest-frame UV absolute magnitude. The COSMOS-based LF of \citet{Meyer_2026_COSMOS}, converted from [O\,III] is shown only for comparison since it samples the same COSMOS field and is not an independent external constraint. We retain this joint fit as our fiducial result and test its dependence on the external data and an assumed shallower slope in Appendix~\ref{app:Hb_sensitivity}.
\item \textit{Br$\alpha$: our data plus an $L^\ast$ prior.} The small low-redshift volume gives limited leverage on the rare bright sources that determine the knee. We use the prior
$\log_{10}\!\left[L^\ast_{\rm H\alpha}/(\mathrm{erg\,s^{-1}})\right]\sim\mathcal{N}(42.15,0.08^2)$ from the joint SHELS and narrow-band H$\alpha$ LF fit of \citet{westra_evolution_2010}. For Br$\alpha$, the corresponding prior center is $\log_{10}L^\ast_{\rm Br\alpha}=40.60$. This Gaussian prior allows the knee to vary and propagates its uncertainty into $\phi^\ast$, $\alpha$, and the CSFRD.
\end{enumerate}

For all lines, the numerical sampling bounds are $-7<\log_{10}(\phi^\ast/\mathrm{Mpc}^{-3})<-1$, $39<\log_{10}[L^\ast_{\rm H\alpha}/(\mathrm{erg\,s^{-1}})]<44$, and $-2.5<\alpha<-0.3$, with uniform density inside these bounds apart from the Br$\alpha$ Gaussian factor. We use \texttt{emcee} \citep{Foreman-Mackey_2013} for the fitting and report posterior medians and 16th--84th percentile intervals in Table~\ref{tab:schechter_parameters}.

\begin{figure*}[!t]
    \centering
    \includegraphics[width=0.88\textwidth]{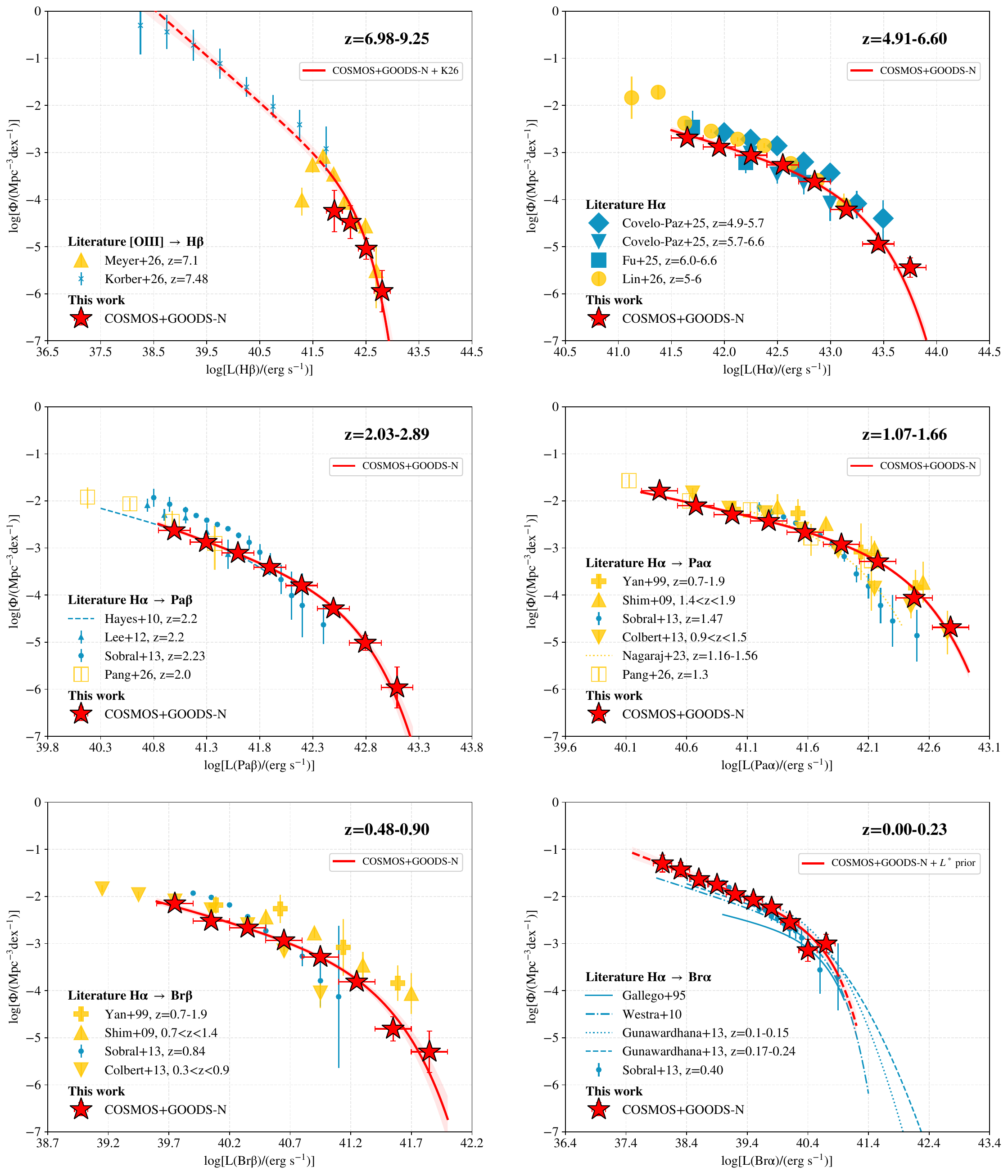}

    \caption{Dust-corrected LFs of the six hydrogen recombination lines from COSMOS+GOODS-N (red stars), compared with literature measurements over the indicated redshift intervals.
Horizontal error bars mark the 0.3-dex bin widths; vertical errors include Poisson noise, cosmic variance, and finite-injection completeness uncertainty. Literature luminosities are converted to those of the emission line indicated on each axis. Yellow denotes measurements without dust correction in the original analysis; open markers identify comparisons requiring an [N\,II] correction, and small symbols denote narrow-band measurements. The red curves and shading show the posterior-median LF and pointwise 16th--84th percentile interval. Solid curves mark the observed luminosity ranges; dashed extensions are shown only for H$\beta$ and Br$\alpha$, which use the K26 faint-end LF and the Gaussian $L^\ast$ prior, respectively (\S\ref{res:line_lfs}). Other literature points are shown for comparison only.
    }
    \label{fig:LF_comparison}
\end{figure*}

For the LF comparison in Figure~\ref{fig:LF_comparison}, solid curves span the luminosity range sampled by our data, defined by the outer edges of the populated bins. For H$\beta$ and Br$\alpha$, dashed curves additionally show the extrapolated fits, while the other four curves are restricted to the observed range.
Overall, the measured dust-corrected LFs show a clear evolution with redshift, which is further investigated through the evolution of the fitted Schechter parameters in \S\ref{disc:para_evo} and Figure~\ref{fig:compare_LF_pare}.
We compare our measurements with previous H$\alpha$ or  [O\,III]+H$\beta$ LFs derived from photometric measurements (\citealp{hayes_h-alpha_2010,lee_dual-narrowband_2012,sobral_large_2013}; K26), slit or fiber spectroscopy \citep{westra_evolution_2010,gunawardhana_galaxy_2013}, and slitless spectroscopic observations \citep{gallego_current_1995,yan_h_1999,shim_global_2009,colbert_predicting_2013,nagaraj_h_2023,Meyer_2026_COSMOS,covelo-paz_h_2025,fu_medium-band_2025,lin_luminosity_2025,pang_glass-jwst_2026}.
Literature H$\alpha$ LFs are converted using the adopted line ratios; the H$\beta$ panel also includes [O\,III]- or blended-line-based conversions. For the displayed $z<4$ comparisons without dust correction we apply $A_{\rm H\alpha}=1$ mag; where needed, the adopted [N\,II] correction is [N\,II]/([N\,II]+H$\alpha$)$=0.25$. 
The H$\alpha$ LF at $z\simeq5$--6 is consistent with recent JWST slitless-spectroscopic measurements in smaller survey areas and extends to higher luminosities.
The lower-redshift Pa$\alpha$, Pa$\beta$, Br$\alpha$, and Br$\beta$ luminosity functions are consistent with the converted H$\alpha$ luminosity functions from previous studies.
Because these longer-wavelength lines are less affected by dust attenuation than H$\alpha$, they provide a useful complement to traditional H$\alpha$ luminosity-function studies.

\begin{table}[!t]
\centering
\footnotesize
\caption{Dust-corrected COSMOS+GOODS-N Luminosity Functions}
\label{tab:dustcorr_LF}
\setlength{\tabcolsep}{3pt}
\renewcommand{\arraystretch}{0.82}
\begin{tabular}{cccc}
\hline\hline
$\log_{10}(L_{\mathrm{line}}/\mathrm{erg\,s^{-1}})$ & $\log_{10}(\Phi/\mathrm{Mpc^{-3}\,dex^{-1}})$ & $N_{\rm CS}$ & $N_{\rm GN}$ \\
\hline
\multicolumn{4}{c}{\hb} \\
\hline
$41.95\pm0.15$ & $-4.24\pm0.44$ & $0$ & $1$ \\
$42.25\pm0.15$ & $-4.48\pm0.35$ & $0$ & $2$ \\
$42.55\pm0.15$ & $-5.05\pm0.22$ & $3$ & $2$ \\
$42.85\pm0.15$ & $-5.95\pm0.44$ & $0$ & $1$ \\
\hline
\multicolumn{4}{c}{\ha} \\
\hline
$41.65\pm0.15$ & $-2.69\pm0.10$ & $33$ & $39$ \\
$41.95\pm0.15$ & $-2.88\pm0.06$ & $189$ & $118$ \\
$42.25\pm0.15$ & $-3.06\pm0.06$ & $548$ & $114$ \\
$42.55\pm0.15$ & $-3.27\pm0.06$ & $580$ & $65$ \\
$42.85\pm0.15$ & $-3.62\pm0.06$ & $290$ & $32$ \\
$43.15\pm0.15$ & $-4.21\pm0.07$ & $74$ & $10$ \\
$43.45\pm0.15$ & $-4.94\pm0.12$ & $14$ & $2$ \\
$43.75\pm0.15$ & $-5.44\pm0.20$ & $4$ & $1$ \\
\hline
\multicolumn{4}{c}{\pab} \\
\hline
$40.95\pm0.15$ & $-2.63\pm0.07$ & $35$ & $27$ \\
$41.25\pm0.15$ & $-2.87\pm0.05$ & $145$ & $44$ \\
$41.55\pm0.15$ & $-3.11\pm0.04$ & $235$ & $55$ \\
$41.85\pm0.15$ & $-3.41\pm0.05$ & $204$ & $26$ \\
$42.15\pm0.15$ & $-3.80\pm0.05$ & $108$ & $11$ \\
$42.45\pm0.15$ & $-4.29\pm0.08$ & $43$ & $0$ \\
$42.75\pm0.15$ & $-5.02\pm0.16$ & $6$ & $2$ \\
$43.05\pm0.15$ & $-5.96\pm0.44$ & $1$ & $0$ \\
\hline
\multicolumn{4}{c}{\paa} \\
\hline
$40.35\pm0.15$ & $-1.79\pm0.05$ & $174$ & $58$ \\
$40.65\pm0.15$ & $-2.10\pm0.04$ & $480$ & $129$ \\
$40.95\pm0.15$ & $-2.29\pm0.04$ & $979$ & $146$ \\
$41.25\pm0.15$ & $-2.43\pm0.04$ & $1271$ & $122$ \\
$41.55\pm0.15$ & $-2.67\pm0.04$ & $960$ & $65$ \\
$41.85\pm0.15$ & $-2.92\pm0.04$ & $579$ & $32$ \\
$42.15\pm0.15$ & $-3.29\pm0.04$ & $259$ & $13$ \\
$42.45\pm0.15$ & $-4.06\pm0.07$ & $45$ & $1$ \\
$42.75\pm0.15$ & $-4.69\pm0.14$ & $10$ & $1$ \\
\hline
\multicolumn{4}{c}{\brb} \\
\hline
$39.75\pm0.15$ & $-2.15\pm0.08$ & $25$ & $22$ \\
$40.05\pm0.15$ & $-2.52\pm0.06$ & $53$ & $29$ \\
$40.35\pm0.15$ & $-2.67\pm0.05$ & $133$ & $33$ \\
$40.65\pm0.15$ & $-2.93\pm0.05$ & $134$ & $23$ \\
$40.95\pm0.15$ & $-3.29\pm0.06$ & $81$ & $8$ \\
$41.25\pm0.15$ & $-3.81\pm0.09$ & $26$ & $3$ \\
$41.55\pm0.15$ & $-4.81\pm0.25$ & $3$ & $0$ \\
$41.85\pm0.15$ & $-5.30\pm0.44$ & $1$ & $0$ \\
\hline
\multicolumn{4}{c}{\bra} \\
\hline
$38.05\pm0.15$ & $-1.30\pm0.18$ & $4$ & $4$ \\
$38.35\pm0.15$ & $-1.43\pm0.15$ & $10$ & $3$ \\
$38.65\pm0.15$ & $-1.64\pm0.13$ & $17$ & $3$ \\
$38.95\pm0.15$ & $-1.75\pm0.12$ & $19$ & $8$ \\
$39.25\pm0.15$ & $-1.96\pm0.12$ & $25$ & $5$ \\
$39.55\pm0.15$ & $-2.08\pm0.12$ & $31$ & $2$ \\
$39.85\pm0.15$ & $-2.24\pm0.12$ & $27$ & $1$ \\
$40.15\pm0.15$ & $-2.55\pm0.14$ & $15$ & $0$ \\
$40.45\pm0.15$ & $-3.15\pm0.23$ & $4$ & $0$ \\
$40.75\pm0.15$ & $-3.00\pm0.20$ & $6$ & $0$ \\
\hline
\hline
\end{tabular}
\par\smallskip
\begin{minipage}{\columnwidth}
\scriptsize
\textit{Note.---}Luminosities refer to the indicated emission line and are expressed in erg s$^{-1}$; $\Phi$ is in Mpc$^{-3}$ dex$^{-1}$. The luminosity interval is the bin center plus/minus half the 0.3-dex width, not a measurement error. The quoted log-LF errors include weighted Poisson noise, field cosmic variance, and finite-injection completeness uncertainty. $N_{\rm CS}$ and $N_{\rm GN}$ count COSMOS and GOODS-N sources, respectively, after $C_{\rm eff}>0.01$; their sum is the total count in each bin. Literature bins are not included in this table.
\end{minipage}
\end{table}

\subsection{Star formation rate functions}
\label{res:line_sfrfs}

\begin{figure*}[!t]
    \centering
    \includegraphics[width=\textwidth]{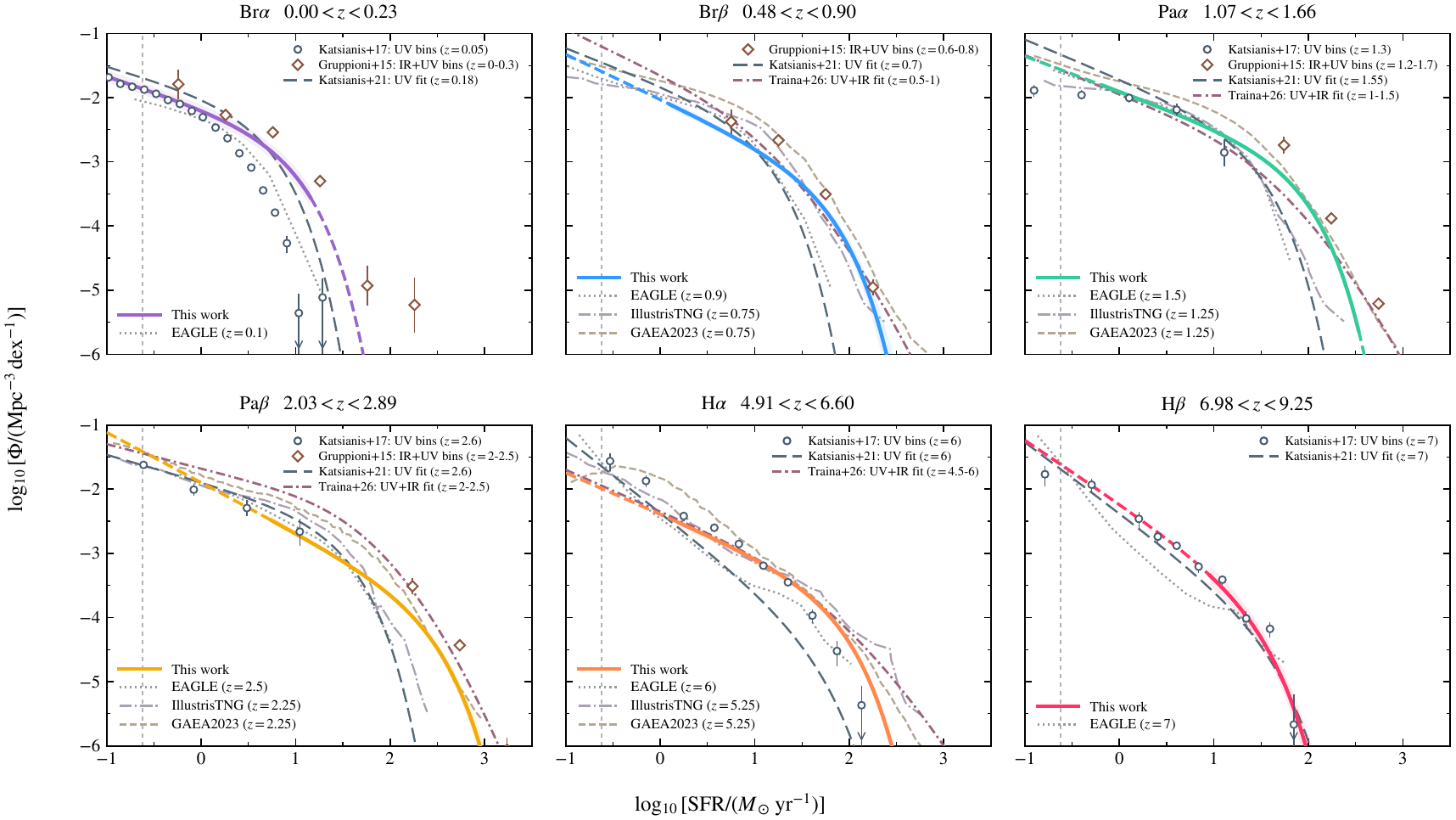}
    \caption{SFRFs inferred from the six COSMOS+GOODS-N line samples, compared with EAGLE (dotted), IllustrisTNG (dash-dotted), and GAEA2023 (dashed) predictions, as well as UV- and IR-based observational determinations.
Panels are arranged in order of increasing redshift from left to right and top to bottom, with common axis limits. Colored curves show our posterior-median Schechter SFRFs, using the same line colors as the CSFRD measurements. Solid segments and shaded pointwise 16th--84th percentile bands indicate the SFR ranges covered by our samples; same-color dashed segments show extrapolations beyond these ranges. For each model, we show an available snapshot near the representative redshift of the corresponding sample, with its actual redshift given in the legend. Only EAGLE is shown for Br$\alpha$ and H$\beta$, as the IllustrisTNG and GAEA2023 comparison data used here do not cover these redshifts. Gray vertical dashed lines mark the common CSFRD integration limit of $0.24~M_\odot\,\mathrm{yr}^{-1}$.
We also show UV- and IR-based SFRFs from \citet{Katsianis2017}, \citet{Gruppioni2015}, \citet{Katsianis2021}, and \citet{Traina_2026} for comparison. Each upper-right legend gives the redshift of the corresponding literature comparison.
    }
    \label{fig:SFR_func}
\end{figure*}

We convert the dust-corrected recombination-line luminosities into SFRs using the equivalent H$\alpha$ luminosities derived in \S\ref{res:line_lfs}.
The SFR is estimated from the H$\alpha$ luminosity following
\begin{equation}
{\rm SFR}\,(M_\odot\,{\rm yr}^{-1})
=
K_{\rm H\alpha}\,
L_{\rm H\alpha},
\end{equation}
where $K_{\rm H\alpha}\simeq10^{-41.30} M_\odot\ \mathrm{yr}^{-1} \mathrm{erg}^{-1}\ \mathrm{s}$ is the H$\alpha$--SFR conversion factor adopted from \citet{Kennicutt_2012} multiplied by a constant factor 0.94 to consider a \citet{Chabrier_2003} IMF \citep{Madau_2014}.
The equivalent H$\alpha$ luminosities inferred from H$\beta$, Pa$\beta$, Pa$\alpha$, Br$\beta$, and Br$\alpha$ are converted in the same way.
The star formation rate functions (SFRFs) can then be directly converted from the dust-corrected line LFs.
Figure~\ref{fig:SFR_func} shows the SFRFs in separate panels for the six line samples, compared with UV- and IR-based observational determinations and predictions from cosmological hydrodynamical simulations and semi-analytic models (SAMs). Solid curves cover the SFR ranges probed by our samples, while dashed extensions show the fitted Schechter functions outside those ranges.
We include the UV SFRF bins compiled by \citet{Katsianis2017}, the IR+UV SFRF bins of \citet{Gruppioni2015}, the UV fits of \citet{Katsianis2021}, and the combined UV+IR fits evaluated from the modified-Schechter parameters in Table~1 of \citet{Traina_2026}. These comparisons use a Chabrier IMF. We approximately rescale luminosity-derived SFRs and number densities to our cosmology at the quoted redshifts, while retaining each study's SFR calibration. They are shown for comparison only and are not included in our fits.  The two Katsianis compilations share some underlying UV measurements.

\subsection{Cosmic star formation rate density}
\label{res:csfrd}

We estimate the CSFRD by integrating the Schechter SFRFs:
\begin{equation}
\rho_{\rm SFR}
=
\int_{{\rm SFR}_{\rm min}}^{{\rm SFR}_{\rm max}}
{\rm SFR}\,
\Phi({\rm SFR})\,
d\log{\rm SFR}.
\end{equation}
The integration uses each posterior realization from the 
LF fit adopted for that line. It propagates the LF covariance and, for Br$\alpha$, the Gaussian $L^\ast$ prior. We use a finite upper limit corresponding to $\log_{10}[L_{\rm H\alpha}/(\mathrm{erg\,s^{-1}})]=45$; the same limit is used for the sensitivity tests in Appendix~\ref{app:Hb_sensitivity}.

For each MCMC realization, we compute the integrated CSFRD and adopt the median and the 16th--84th percentile range as the final value and uncertainty. 
For consistency with previous works, we set the lower integration limit to $0.24\ M_\odot\ \mathrm{yr}^{-1}$, labeled in Figure~\ref{fig:SFR_func}, corresponding to an absolute UV magnitude limit of $M_{\mathrm{UV}}=-17$ mag or UV luminosity $0.03\,L^*_{\mathrm{UV},z=3}$  \citep[e.g.,][]{Bouwens_2015,Adams_2024,Willott_2024}. We adopt this limit because galaxies are commonly observed at $z \geq 6$ down to magnitudes of $M_{\mathrm{UV}}=-17$ mag, and here we use the same lower SFR limit to fairly compare with the UV- and \ha-based CSFRD estimates across the full redshift range.

\begin{figure*}[!t]
    \centering
    \includegraphics[width=\textwidth]{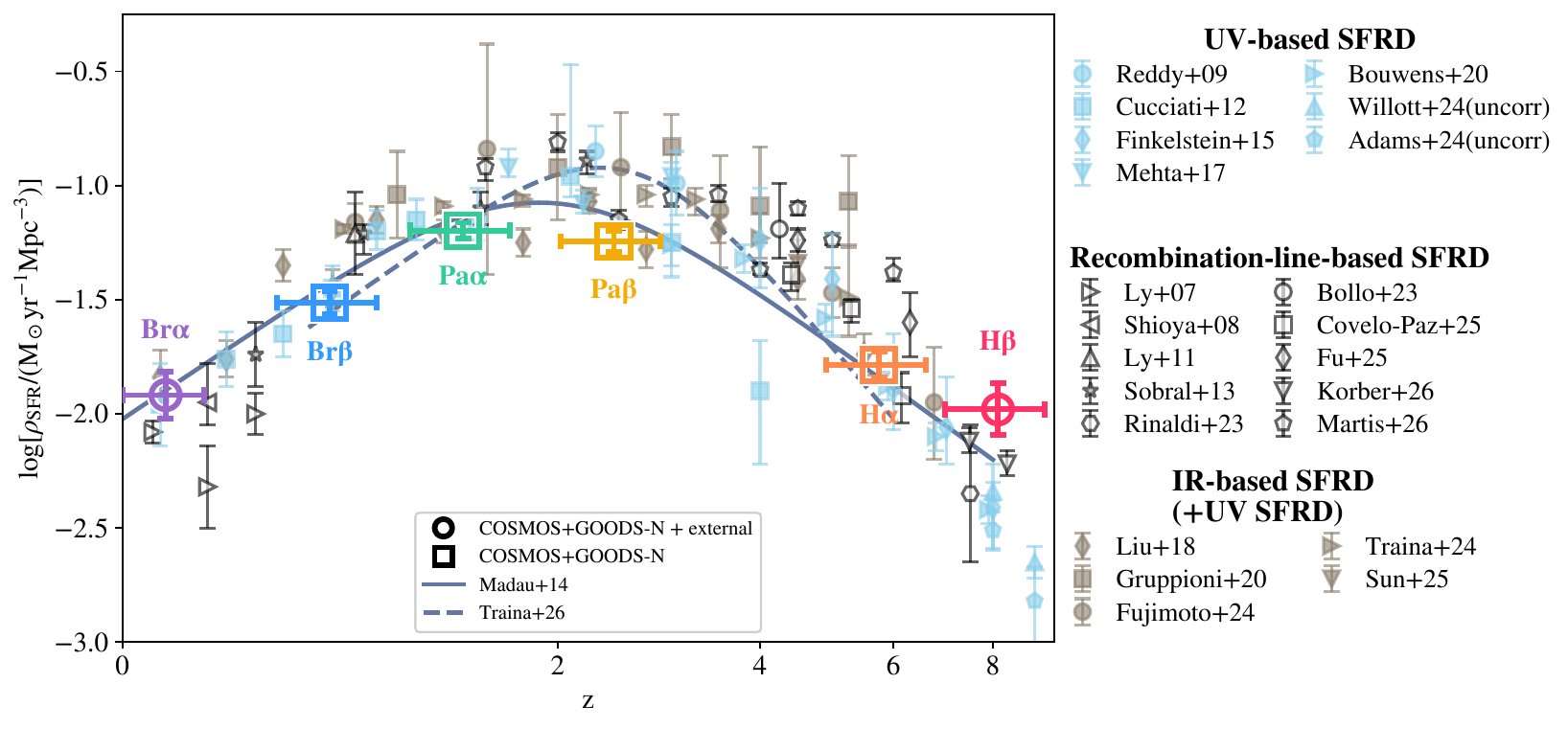}
    \caption{CSFRD inferred from the Schechter fit to each COSMOS+GOODS-N recombination-line sample, integrated above $0.24~M_\odot\,\mathrm{yr}^{-1}$.
Colored open squares denote H$\alpha$, Pa$\alpha$, Pa$\beta$, and Br$\beta$ results fitted only to our data. Open circles denote H$\beta$ with the K26 faint-end LF and Br$\alpha$ with the Gaussian $L^\ast$ prior. Vertical bars are posterior 16th--84th percentile intervals, and horizontal bars show each line's redshift interval. The H$\beta$ interval is conditional on the adopted faint-end constraint; alternative slope assumptions are examined in Appendix~\ref{app:Hb_sensitivity}. Blue-gray solid and dashed curves give the \citet{Madau_2014} and \citet{Traina_2026} histories, respectively; comparisons with galaxy-formation models are presented in Figure~\ref{fig:CSFRD_models}. Gray symbols show literature recombination-line-based estimates, while light-blue and brown symbols show UV- and IR-based estimates, respectively; the two uncorrected UV datasets are identified in the legend. 
    }
    \label{fig:CSFR_evo}
\end{figure*}

The CSFRD measurements derived from the different recombination-line samples are shown in Figure~\ref{fig:CSFR_evo}.
We compare our results with previous recombination-line-based estimates (\citealp{Ly_2007, Shioya_2008,Ly_2011,sobral_large_2013, Bollo_2023, Rinaldi_2023, covelo-paz_h_2025,fu_medium-band_2025}; K26; \citealp{Martis_2026_medband_HaLF}), IR-based estimates \citep{Liu_2018,gruppioni_alpine-alma_2020,fujimoto_alma_2024,Traina_2024,sun_2025}, and UV-based estimates \citep{Reddy_2009,Cucciati_2012,Finkelstein_2015,Mehta_2017,Bouwens_2020,Adams_2024}.

Overall, our measurements recover the well-known rise of the CSFRD from the local Universe toward a broad maximum around cosmic noon, followed by a decline toward higher redshift. Our results at $z\lesssim6$ are broadly consistent with the cosmic star formation history of \citet{Madau_2014}, after conversion to a common IMF scale. The fiducial H$\beta$ estimate lies above this reference history at its redshift, but only about $18\%$ of its integral lies above our spectroscopic luminosity limit.  Appendix~\ref{app:Hb_sensitivity} quantifies how the inferred CSFRD changes with the faint-end assumptions. The recombination lines probe shorter star-formation timescales and are more sensitive to recent star formation, which can yield higher inferred SFRs for galaxies with rapidly rising or bursty star formation histories, as commonly expected at high redshift \citep{Smit_2016,Faisst_2019,Endsley_2024}. However, the current comparison alone does not isolate this effect from LF extrapolation and external-constraint systematics.


\begin{deluxetable*}{lccccccc}[!t]
\tablewidth{0pt}
\tabletypesize{\scriptsize}
\tablecaption{Schechter Parameters and Derived CSFRDs\label{tab:schechter_parameters}}
\tablehead{
\colhead{Line} & \colhead{Redshift} & \colhead{$\log(\phi^\ast/\mathrm{Mpc}^{-3})$} & \colhead{$\log[L^\ast/(\mathrm{erg\,s^{-1}})]$} & \colhead{$\alpha$} & \colhead{$\log\rho_{\rm SFR}$\tablenotemark{a}} & \colhead{$\log L$ range$^{b}$} & \colhead{Additional data / constraints}
}
\startdata
\bra & $0.00$--$0.23$ & $-2.94_{-0.14}^{+0.14}$ & $40.62_{-0.07}^{+0.07}$ & $-1.49_{-0.05}^{+0.06}$ & $-1.92_{-0.10}^{+0.10}$ & $37.90$--$40.90$ & $L^\ast$ Gaussian prior \\
\brb & $0.48$--$0.90$ & $-3.55_{-0.16}^{+0.14}$ & $41.17_{-0.08}^{+0.10}$ & $-1.70_{-0.08}^{+0.09}$ & $-1.51_{-0.05}^{+0.05}$ & $39.60$--$42.00$ & \nodata \\
\paa & $1.07$--$1.66$ & $-3.25_{-0.06}^{+0.06}$ & $42.17_{-0.03}^{+0.03}$ & $-1.55_{-0.02}^{+0.02}$ & $-1.20_{-0.03}^{+0.03}$ & $40.20$--$42.90$ & \nodata \\
\pab & $2.03$--$2.89$ & $-4.08_{-0.14}^{+0.13}$ & $42.39_{-0.07}^{+0.08}$ & $-1.78_{-0.07}^{+0.07}$ & $-1.24_{-0.05}^{+0.05}$ & $40.80$--$43.20$ & \nodata \\
\ha & $4.91$--$6.60$ & $-3.89_{-0.10}^{+0.10}$ & $43.07_{-0.04}^{+0.05}$ & $-1.64_{-0.06}^{+0.06}$ & $-1.79_{-0.06}^{+0.06}$ & $41.50$--$43.90$ & \nodata \\
\hb & $6.98$--$9.25$ & $-3.92_{-0.34}^{+0.30}$ & $42.19_{-0.11}^{+0.13}$ & $-1.99_{-0.11}^{+0.11}$ & $-1.98_{-0.11}^{+0.12}$ & $38.04$--$43.00$ & K26 \\
\enddata
\tablenotetext{a}{CSFRD in $M_\odot\,\mathrm{yr}^{-1}\,\mathrm{Mpc}^{-3}$, integrated above $0.24~M_\odot\,\mathrm{yr}^{-1}$.}
\tablenotetext{b}{Luminosity interval in erg s$^{-1}$ spanned by the fitted bins for the emission line listed in each row, including the K26 bins for H$\beta$. The H$\beta$ bins from this work alone cover $41.80\leq\log_{10}L\leq43.00$.}
\tablecomments{Errors are posterior 16th--84th percentile intervals.}
\end{deluxetable*}

\section{Discussion} \label{sec:discuss}

\subsection{Luminosity and Density Evolution of the Star-forming Galaxy Population} \label{disc:para_evo}

\begin{figure}[!t]
    \centering
    \includegraphics[width=\columnwidth]{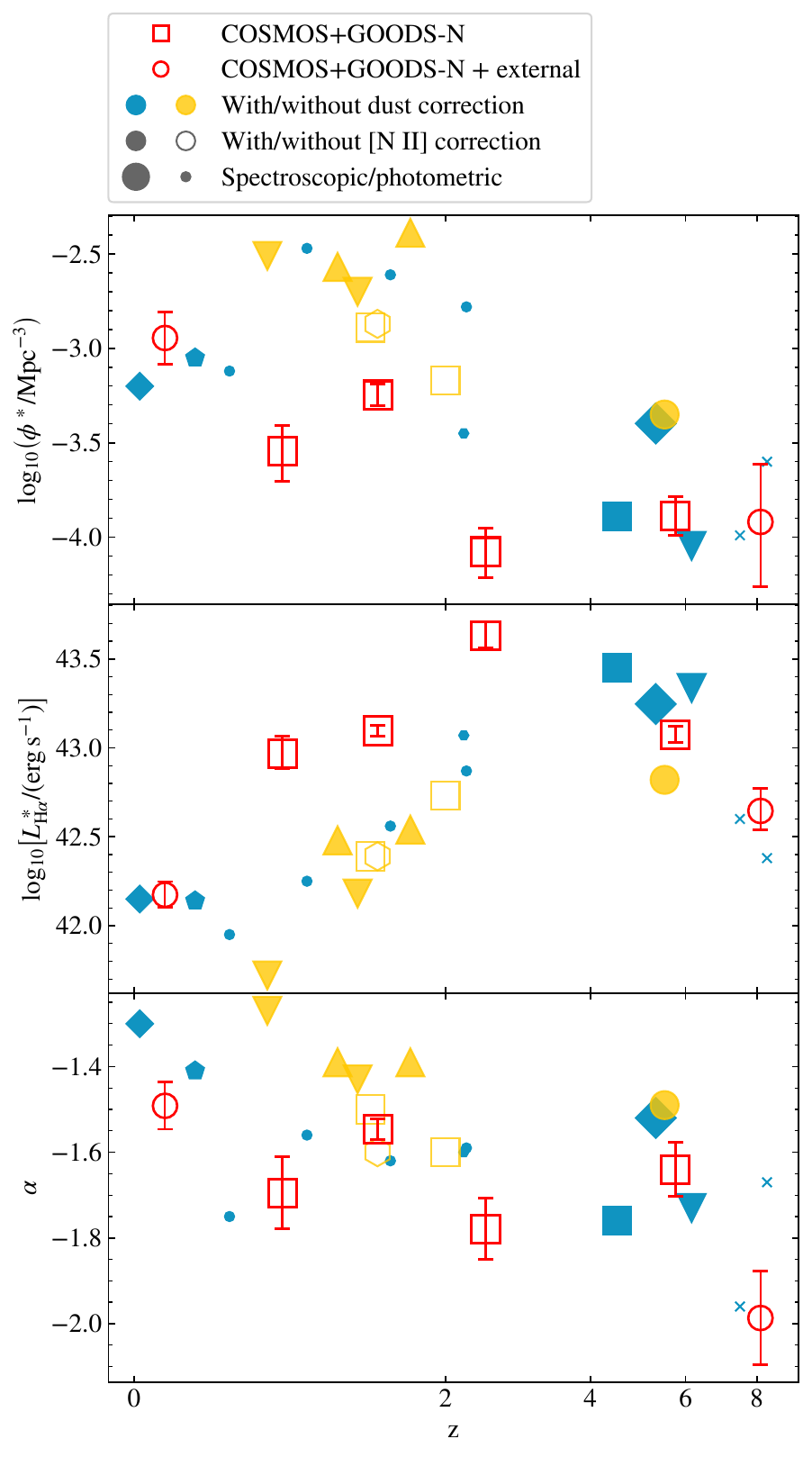}
    \caption{Evolution of the Schechter parameters: $\log_{10}(\phi^\ast/\mathrm{Mpc}^{-3})$ (top), H$\alpha$-equivalent $\log_{10}[L^\ast_{\rm H\alpha}/(\mathrm{erg\,s^{-1}})]$ (middle), and $\alpha$ (bottom).
Red open squares show the four COSMOS+GOODS-N fits using only our bins; red open circles show H$\beta$ with K26 and Br$\alpha$ with the Gaussian $L^\ast$ prior. Error bars are posterior 16th--84th percentile intervals. Literature datasets shown as points in Figure~\ref{fig:LF_comparison} use the same marker shapes here. For the literature points, blue/yellow indicate results reported with/without dust correction, and filled/open symbols indicate results with/without [N\,II] correction where applicable. Larger symbols represent spectroscopic measurements; smaller symbols represent photometric measurements, including narrow-band surveys and K26. No additional dust correction is applied to the literature parameter compilation in this figure. Unlike Table~\ref{tab:schechter_parameters}, the middle panel uses H$\alpha$-equivalent luminosities rather than luminosities of the corresponding emission lines.
    }
    \label{fig:compare_LF_pare}
\end{figure}

The evolution of the Schechter parameters provides a compact view of how the star-forming galaxy population changes across cosmic time.
Figure~\ref{fig:compare_LF_pare} shows the redshift evolution of the normalization $\phi^\ast$, the \ha-equivalent characteristic luminosity $L^\ast_{\rm H\alpha}$, and the faint-end slope $\alpha$.
From the local Universe to cosmic noon, the most prominent trend is the strong increase in $L^\ast_{\rm H\alpha}$.
This is consistent with previous \ha\ LF measurements at $z<3$. For example, \citet{sobral_large_2013} found a rapid increase in the characteristic \ha\ luminosity from $z=0.4$ to $z=2.23$, while $\alpha$ remained approximately constant.
Across the redshift range probed, our $\phi^\ast$ generally decreases toward higher redshift.
The rise of the cosmic star formation activity toward $z\sim2$ is therefore associated primarily with increasing characteristic luminosities, rather than with an increasing LF normalization.

The evolution of $L^\ast_{\rm H\alpha}$ can be related to the growth of dark-matter halos.
At fixed halo mass, the mean halo mass-accretion rate increases rapidly with redshift \citep{Fakhouri_2010}.
Empirical galaxy--halo models further suggest that the conversion of accreted baryons into stars is most efficient in halos around $M_{\rm h}\sim10^{12}\,M_\odot$, with a weaker redshift dependence than that of the halo accretion rate \citep{behroozi_average_2013}.
The evolution of $L^\ast_{\rm H\alpha}$ may therefore partly reflect changes in the cosmological gas supply, with halo abundance and accretion contributing to the CSFRD peak at $z\sim2$--3 \citep{Harikane_2022}. 

 The H$\beta$ joint fit gives $\alpha\simeq-2$ at $z\sim8$, conditional on the K26 measurements and their empirical [O\,III]/H$\beta$ conversion. Our H$\beta$ bins alone also admit a shallower $\alpha=-1.6$ fit (Appendix~\ref{app:Hb_sensitivity}). For a slope near $-2$ and $L\ll L^\ast$, galaxies contribute approximately equally to the luminosity density per logarithmic luminosity interval. The resulting sensitivity to the integration limit makes direct faint-end measurements particularly important at these redshifts.

\subsection{Comparison with galaxy formation models}
\label{disc:sim_comparison}

The observational comparisons in Figure~\ref{fig:SFR_func} help distinguish differences in the SFR distribution from differences in its integral. In several redshift intervals, the IR+UV determinations have a more extended high-SFR tail than the UV-based fits, while the two approaches also differ around and below the knee \citep{Katsianis2017,Gruppioni2015,Katsianis2021,Traina_2026}. Our recombination-line SFRFs provide a complementary comparison over the solid segments constrained by our samples. Differences among these determinations can reflect selection, dust treatment, SFR calibration, and the non-identical redshift intervals; comparisons outside our measured SFR ranges additionally depend on extrapolation.

The broad redshift coverage of our sample allows us to compare the observed SFRFs and CSFRD with predictions from different galaxy formation models. 
In Figure~\ref{fig:SFR_func}, we compare our SFRFs with the EAGLE hydrodynamical simulation \citep{Schaye_2015,Crain2015,Katsianis2017}, the IllustrisTNG simulation \citep{Pillepich2018}, and the GAEA2023 semi-analytic model \citep{DeLucia2024}. 
These models adopt different numerical methods and feedback prescriptions, and therefore provide complementary tests of galaxy formation physics.
Overall, all three models reproduce an evolutionary trend with redshift that is broadly similar to that inferred from our observations.

At redshifts below cosmic noon ($z\lesssim2$), EAGLE broadly reproduces the observed SFRFs around and below the knee, but declines more rapidly than our measurements toward the high-SFR end. 
This indicates that EAGLE does not produce enough intensely star-forming galaxies, especially at low to intermediate redshifts. 
A similar deficit has been reported in previous comparisons with UV- and IR-based SFRFs \citep[e.g.,][]{Katsianis2017,Traina_2026}. 
The discrepancy may be related to the adopted AGN-feedback prescriptions. 
The relatively small box size of the reference EAGLE simulation ($\sim100$~cMpc) may also limit the sampling of rare high-SFR galaxies.
IllustrisTNG extends to substantially higher SFRs and predicts higher number densities above the SFRF knee than EAGLE. 
It therefore provides a better description of our measurements in the high-SFR regime. 
The difference may partly arise from the different implementations of stellar and AGN feedback. 
In particular, IllustrisTNG includes an updated galactic-wind model and a kinetic AGN-feedback mode at low black-hole accretion rates. 
GAEA2023 provides a complementary comparison because it is a semi-analytic model built on dark-matter halo merger trees from large cosmological $N$-body simulations. 
It includes an updated treatment of black-hole growth and AGN feedback, including explicit quasar-driven outflows \citep{DeLucia2024}.
Its much larger statistical volume enables better sampling of rare galaxies than typical high-resolution hydrodynamical simulations. 
GAEA2023 reproduces the observed bright end relatively well over much of the redshift range explored here. 
The comparison therefore constrains two different regimes, the bright end is sensitive to rare populations and the regulation of star formation in massive galaxies, while the faint end provides a test of star formation and stellar feedback in abundant low-mass systems.

To connect the SFRF comparison to the integrated star-formation budget, we compare our CSFRDs with integrals of the available model SFRFs in Appendix~\ref{app:CSFRD_models} (Figure~\ref{fig:CSFRD_models}). We integrate only over the supplied SFR ranges, without extrapolating either end. Within these ranges, GAEA2023 generally yields higher SFR densities than our measurements, whereas IllustrisTNG is closer and EAGLE tends to yield lower values. 

Figure~\ref{fig:CSFRD_models} also includes the COLIBRE L200m6 hydrodynamical simulation \citep{Schaye_2026_COLIBRE,Chaikin2026}, whose feedback parameters are calibrated primarily against low-redshift galaxy properties rather than the cosmic star formation history. Unlike the SFRF integrals, this prediction sums the SFRs of galaxies with $M_\ast\geq10^7~M_\odot$ within three-dimensional apertures of $50$~kpc. It closely follows our Brackett, Paschen, and H$\alpha$ measurements through cosmic noon and to $z\sim6$, while the fiducial H$\beta$ estimate lies above it at $z\sim8$.  The H$\beta$ offset depends on the faint-end assumptions tested in Appendix~\ref{app:Hb_sensitivity}. 

Overall, the SFRF and CSFRD provide complementary constraints on galaxy formation models. 
A model may reproduce the integrated CSFRD while still showing differences in the predicted distribution of galaxies as a function of SFR. For example, a deficit of rare high-SFR galaxies may have only a limited impact on the total CSFRD if most of the star formation is contributed by galaxies around the knee of the SFRF.

\subsection{Potential Systematic Uncertainties and Environmental Effects}  \label{disc:bias}

Several systematic uncertainties should be considered when interpreting the CSFRD evolution derived in this work. First, we adopt Case B recombination when converting different hydrogen lines to an equivalent \ha\ luminosity. The dust-corrected line ratios in our calibration sample support this approximation (\S\ref{method:sfr_tracers}). The narrow bands indicate that temperature and density variations within Case B produce smaller changes in the line ratios than the observed scatter. However, they do not quantify the possible effects of departures from Case B. Continuum pumping and finite Lyman-line optical depths can alter the intrinsic ratios \citep{Luridiana_2009,mcclymont_density-bounded_2025}, biasing both the conversion to \ha-equivalent luminosities and dust corrections inferred from Balmer decrements.

Separately, the conversion from recombination-line luminosity to SFR assumes that most stellar ionizing photons ionize the nebular gas. Ionizing-photon escape or absorption by dust before hydrogen ionization reduces the line luminosity at fixed SFR and can therefore lead to an underestimate of the CSFRD, even if the line ratios remain close to Case B. Direct measurements of Lyman-continuum escape are generally inaccessible during reionization, leaving indirect diagnostics and lower-redshift analogs as important constraints \citep{nakajima_ionization_2014,izotov_low-redshift_2018,chisholm_far-ultraviolet_2022,flury_low-redshift_2022}.

The adopted H$\alpha$--SFR conversion introduces another uncertainty. 
The \citet{Kennicutt_2012} calibration assumes approximately solar metallicity, whereas low-mass galaxies at high redshift are typically more metal poor. 
Lower-metallicity stellar populations produce more ionizing photons per unit SFR and hence larger H$\alpha$ luminosities at fixed SFR. 
For example, theoretical stellar population models suggest that the \ha--SFR conversion may vary with metallicity. \citet{Clarke_2024}, based on BPASS models, adopt a conversion coefficient of 41.59, compared with $\simeq41.37$ for more metal-rich populations.
We adopt a single conversion throughout this work to maintain a homogeneous SFR scale across all six recombination-line samples.

Dust attenuation may introduce additional tracer-dependent offsets because the six lines span a wide range in rest-frame wavelength. 
Recent JWST observations also suggest that nebular attenuation curves at high redshift may differ from commonly adopted low-redshift curves and show object-to-object variation \citep[e.g.,][]{Sanders_2025_neb_dust_curve,Reddy_2026}. Patchy dust coverage can flatten the effective attenuation curve \citep{Reddy_2026}. At fixed $A_V$, a flatter curve reduces the differential attenuation between bluer and redder lines. At fixed observed Balmer decrement, however, a flatter curve can require a larger attenuation normalization, so its effect on the corrected luminosities is not determined by the slope alone. Such geometry could contribute to tracer-dependent offsets and scatter. 

We exclude AGN candidates with broad-line-dominated profiles and point-like image morphologies, using the line-profile, compactness, and visual criteria described in \S\ref{data:sample_construction}. Narrow-line AGN or composite systems may remain in the sample, and their nuclear emission could raise the bright end of the recombination-line LFs and the inferred SFRs. Conversely, the host galaxies of excluded AGNs may also contribute to the CSFRD through star formation, which is omitted from our census because its line emission cannot be reliably separated from the nuclear component.

Extrapolation below the observational limit introduces a dependence on the LF shape and, where used, external constraints or parameter priors. The H$\beta$ tests in Appendix~\ref{app:Hb_sensitivity} show that refitting the normalization and knee can change the CSFRD substantially when the faint-end assumptions change; a shallower slope does not necessarily lower the integral if the external LF is retained. The quoted posterior intervals are conditional on the adopted fit and do not encompass this variation between assumptions. Deeper H$\beta$ spectroscopy spanning the knee and faint end is needed to constrain the slope directly and test the photometric line-separation and attenuation corrections.


Finally, we test the sensitivity to known overdensities by excluding full-field redshift slices: $2.08\leq z<2.11$ around ZFIRE \citep{Yuan_2014_cluster} and $2.40\leq z<2.52$ around Hyperion \citep{Cucciati_2018_cluster} from the COSMOS Pa$\beta$ sample, and, in a separate test, $5.17\leq z<5.20$ around HDF850.1 \citep{HerardDemanche_2025_HDF850} from the GOODS-N H$\alpha$ sample. For each test, we subtract the affected field's slices from its effective-volume integrals, retain the other field's redshift coverage and volume, reapply $C_{\rm eff}>0.01$, and recompute the LF covariance before refitting with unchanged luminosity bins. The COSMOS cuts remove 212 sources from the fiducial sample and 17.4\% of that field's geometric volume. The Pa$\beta$ slope steepens from $-1.78$ to $-1.88$, and the CSFRD above $0.24~M_\odot\,\mathrm{yr}^{-1}$ increases by 0.030 dex (7.0\%), despite a $\simeq4\%$ decrease in the fitted contribution above our observed luminosity range's lower edge, owing to greater faint-end extrapolation. The GOODS-N cut removes 92 fiducial sources and 1.9\% of its geometric volume; eight additional sources fall below the recomputed joint completeness threshold. The lowest H$\alpha$ LF bin decreases by $\simeq35\%$, while $\alpha$ flattens from $-1.64$ to $-1.60$ and the integrated CSFRD decreases by 0.042 dex (9.3\%). Thus, these exclusions produce modest changes in the integrated CSFRDs, although individual LF bins can be more sensitive to the sampled environments.

\section{Summary}
\label{sec:summary}

In this work, we use JWST/NIRCam F444W WFSS observations from 
COSMOS-3D and GOODS-N to construct a recombination-line-based view of 
cosmic star formation over $z\sim0$--9. 
We use six hydrogen recombination lines, \bra, \brb, \paa, 
\pab, \ha, and \hb, which enter the same observed wavelength 
window at different redshifts. 
After correcting for line-detection completeness, we construct 
LFs, convert all line luminosities to equivalent 
\ha\ luminosities, derive SFRFs, and measure the 
CSFRD within a common framework. 
Our main results are summarized as follows.

\begin{enumerate}

\item
We empirically test the consistency of the different hydrogen recombination lines as SFR tracers using galaxies with simultaneous Balmer and Paschen or Brackett measurements. After correcting for dust attenuation, the observed \paa/\ha, \pab/\ha, \bra/\ha, and \brb/\ha\ luminosity ratios are broadly consistent with the Case B predictions for $T_e=10^4\,{\rm K}$ and $n_e=100\,{\rm cm}^{-3}$, with absolute median offsets of $0.01$--$0.16$ dex and scatters of $0.15$--$0.24$ dex. The recombination-line SFRs are also closely correlated with the 10 Myr-averaged SED-based SFRs.
\item
The recombination-line LFs show substantial evolution across cosmic time.
The wide area and large survey volume of COSMOS-3D allow us to probe the bright end of the LFs, while the deeper GOODS-N observations extend the measurements to the faint end. 
After converting previous \ha\ LFs at comparable redshifts to the corresponding recombination-line luminosities using the adopted intrinsic line ratios, we find broad agreement over the overlapping luminosity ranges.

\item
Combining Brackett, Paschen, and Balmer recombination lines, we trace the evolution of the CSFRD over $z\sim0$--9. The Brackett and Paschen measurements constrain star formation at low and intermediate redshifts, while the Balmer measurements extend this view to the reionization era. Integrating the SFRFs above $0.24\,M_\odot\,{\rm yr}^{-1}$, we obtain $\log_{10}[\rho_{\rm SFR}/(M_\odot\,{\rm yr}^{-1}\,{\rm Mpc}^{-3})]=-1.92$, $-1.51$, $-1.20$, $-1.24$, $-1.79$, and $-1.98$ for Br$\alpha$, Br$\beta$, Pa$\alpha$, Pa$\beta$, H$\alpha$, and H$\beta$, respectively. Together, they recover the rise of cosmic star formation toward cosmic noon and its subsequent decline toward the present day, providing a complementary view of the cosmic star formation history to UV- and IR-based studies. The fiducial H$\beta$ value is conditional on K26, with about $18\%$ of its inferred CSFRD above our spectroscopic limit. A fixed-$\alpha=-1.6$ fit to our bins alone gives a CSFRD lower by $0.91$ dex; this alternative illustrates the combined sensitivity to the slope assumption and external data (Appendix~\ref{app:Hb_sensitivity}).

\item
Across the six recombination-line samples, the characteristic H$\alpha$-equivalent luminosity increases toward cosmic noon.  The H$\beta$ faint-end slope remains dependent on external photometric constraints, deeper spectroscopy is needed for a direct measurement. The evolution of the characteristic luminosity and LF normalization links changes in the star-forming galaxy population to the cosmic star formation history.

\item
Comparisons with galaxy-formation models show broad agreement in the overall evolution of cosmic star formation, while differences remain in the detailed shape of the SFRFs. This demonstrates that the SFRF provides additional constraints beyond the integrated CSFRD, particularly on the regulation of star formation across different galaxy populations.

\end{enumerate}

Together, these measurements establish a recombination-line-based benchmark for cosmic star formation across $z\sim0$--9 using a common observational configuration and SFR scale.

\section*{Acknowledgment}
We acknowledge support from the National Science Foundation of China (12225301).

\vspace{5mm}
\facilities{JWST (NIRCam, NIRSpec), HST (ACS)}

\software{astropy \citep{astropy_2013},  
          photutils \citep{photutils},
          bagpipes \citep{Carnall_2018},
          emcee \citep{Foreman-Mackey_2013},
          JWST \citep{jwst_bushouse_2023},
          }

\appendix
\section{CSFRD Comparison with Galaxy-Formation Models}
\label{app:CSFRD_models}

\begin{figure}[!t]
    \centering
    \includegraphics[width=\columnwidth]{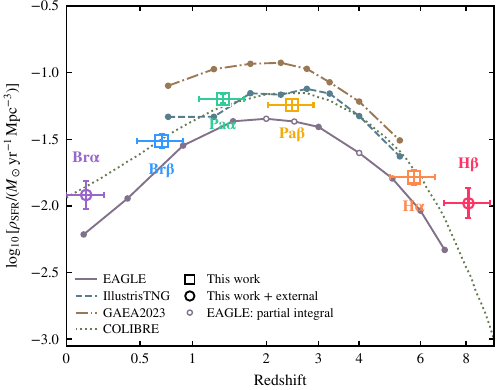}
    \caption{CSFRD comparison with galaxy-formation models. Colored symbols reproduce the measurements in Figure~\ref{fig:CSFR_evo}, with the same colors and marker shapes. The EAGLE, IllustrisTNG, and GAEA2023 curves integrate the available SFRFs from the larger of $0.24~M_\odot\,\mathrm{yr}^{-1}$ and the lower input endpoint to the upper endpoint, without extrapolation. Open EAGLE circles mark incomplete low-SFR coverage; filled circles reach the adopted limit. Their sizes are identical, and connecting lines guide the eye. The olive dotted COLIBRE L200m6 curve instead selects $M_\ast\geq10^7~M_\odot$ galaxies.}
    \label{fig:CSFRD_models}
\end{figure}

Figure~\ref{fig:CSFRD_models} compares our CSFRD measurements with integrals of the sampled EAGLE, IllustrisTNG, and GAEA2023 SFRF curves used in Figure~\ref{fig:SFR_func} \citep{Katsianis2017,Traina_2026}. For each curve, we integrate from the larger of $0.24~M_\odot\,\mathrm{yr}^{-1}$ and its lowest available SFR to its highest available SFR, without extrapolating either end. 
The EAGLE curves at $z=2$, $2.5$, and $4$ begin at approximately $0.58$, $0.59$, and $0.72~M_\odot\,\mathrm{yr}^{-1}$, respectively. We distinguish these partial integrals with open markers of the same size as the filled model markers. For all curves, the bright-end coverage is finite. 
For comparison, we also show the COLIBRE L200m6 CSFRD for galaxies with $M_\ast\geq10^7~M_\odot$, using SFRs within three-dimensional apertures of $50$~kpc \citep{Chaikin2026}. 

\section{Sensitivity of the H$\beta$ LF to Faint-End Assumptions}
\label{app:Hb_sensitivity}

\begin{figure}[!t]
    \centering
    \includegraphics[width=\columnwidth]{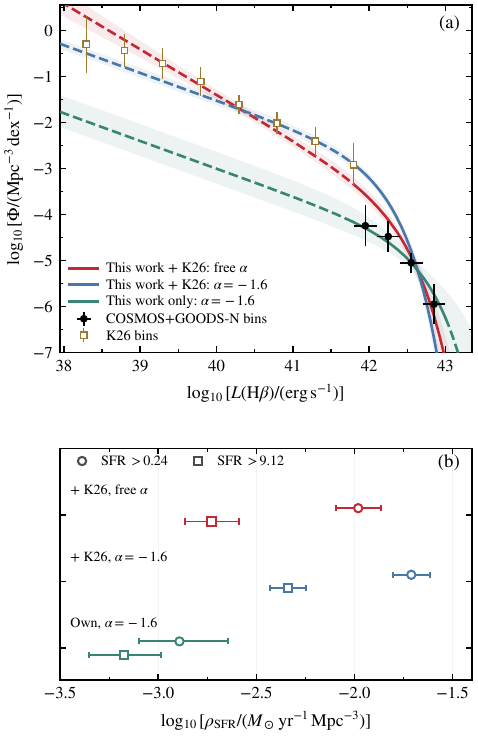}
    \caption{Sensitivity of the H$\beta$ LF and CSFRD to faint-end assumptions. (a) Black circles show the COSMOS+GOODS-N bins and gold squares show K26. Red denotes the fiducial joint fit with free $\alpha$; blue and green denote fixed-$\alpha=-1.6$ fits with and without K26, respectively. Curves and bands show pointwise posterior medians and 16th--84th percentiles. Solid segments span our populated bins; dashed segments extend beyond them. (b) Circles integrate above SFR $=0.24~M_\odot\,\mathrm{yr}^{-1}$ and squares above our sample's lower luminosity edge, corresponding to SFR $=9.12~M_\odot\,\mathrm{yr}^{-1}$. All use the same upper limit as in \S\ref{res:csfrd}. Error bars are posterior 16th--84th percentiles, conditional on each fit's assumptions.}
    \label{fig:Hb_sensitivity}
\end{figure}

Figure~\ref{fig:Hb_sensitivity} compares the fiducial H$\beta$ joint fit with two alternatives fixing $\alpha=-1.6$, with and without K26. Both refit $\phi^\ast$ and $L^\ast$ using the same parameter bounds, binning, and relevant covariance submatrices; all CSFRDs use the integration limits in \S\ref{res:csfrd}.

Retaining K26 while fixing the shallower slope raises the CSFRD by $0.27$ dex: the fit compensates with a higher normalization and lower $L^\ast$, but worsens the agreement with our bins. Without K26, the fixed-slope fit describes our bins and yields a CSFRD $0.91$ dex below the fiducial value. The contrast reflects the covariance among Schechter parameters and the external faint-end constraint, rather than the slope alone.

Only $18\%$ of the fiducial integral lies above our spectroscopic limit. The posterior intervals are conditional on each fit's assumptions, with K26 line-separation and attenuation uncertainties entering through its published LF errors. Deeper H$\beta$ spectroscopy across the knee and faint end would directly constrain $\alpha$ and test these photometric conversions, reducing the dependence of the high-redshift CSFRD on external data and LF extrapolation.

\clearpage
\bibliography{sample631}{}
\bibliographystyle{aasjournal}

\end{document}